\documentclass[3p]{earticle}

\PassOptionsToPackage{hyphens}{url}

\usepackage[utf8]{inputenc} 
\usepackage[T1]{fontenc}    
\usepackage{hyperref}       
\usepackage{url}            
\usepackage{booktabs}       
\usepackage{amsfonts}       
\usepackage{nicefrac}       
\usepackage{microtype}      
\usepackage{lipsum}
\usepackage{graphicx}
\usepackage{amssymb}
\usepackage{amsmath,amsthm}
\usepackage{array}
\usepackage{subfig}
\usepackage{capt-of}
\usepackage{mathtools}
\usepackage{natbib}
\usepackage{xcolor}
\usepackage[title]{appendix}

\usepackage{lmodern}

\biboptions{super}

\begin{document}

\begin{frontmatter}

\title{Persistence of the Omicron variant of SARS-CoV-2 in Australia: \\ The impact of fluctuating social distancing}

\author{Sheryl L. Chang$^{a,b}$, \ Quang Dang Nguyen$^{a}$, \ Alexandra Martiniuk$^{c}$, \ Vitali Sintchenko$^{b,c,d,e}$, \\ Tania C. Sorrell$^{b,c}$, \ Mikhail Prokopenko$^{a,b}$}
\address{$^{a}$  Centre for Complex Systems, Faculty of Engineering, The University of Sydney, Sydney, NSW 2006, Australia\\
$^{b}$  Sydney Institute for Infectious Diseases, The University of Sydney, Westmead, NSW 2145, Australia\\ 
$^{c}$  Faculty of Medicine and Health, The University of Sydney, NSW, 2006, Australia\\
$^{d}$ Centre for Infectious Diseases and Microbiology -- Public Health, Westmead Hospital, Westmead, NSW 2145, Australia\\
$^{e}$ Institute of Clinical Pathology and Medical Research, NSW Health Pathology, Westmead, NSW 2145, Australia
  \ \\
Correspondence to: \textbf{sheryl.chang@sydney.edu.au} 
}

\begin{abstract}
We modelled emergence and spread of the Omicron variant of SARS-CoV-2 in Australia between December 2021 and June 2022. This pandemic stage exhibited a diverse epidemiological profile with emergence of co-circulating sub-lineages of Omicron, further complicated by differences in social distancing behaviour which varied over time. Our study delineated distinct phases of the Omicron-associated pandemic stage, and retrospectively quantified the adoption of social distancing measures, fluctuating over different time periods in response to the observable incidence dynamics. We also modelled the corresponding disease burden, in terms of hospitalisations, intensive care unit occupancy, and mortality. Supported by good agreement between simulated and actual health data, our study revealed that the nonlinear dynamics observed in the daily incidence and disease burden  were determined not only by introduction of sub-lineages of Omicron, but also by the fluctuating adoption of social distancing measures. Our high-resolution  model can be used in design and evaluation of public health interventions during future crises. 
\end{abstract}

\end{frontmatter}

\section{Introduction}
\label{intro}

On 24 November 2021 the B.1.1.529 variant of SARS-CoV-2 was reported to the World Health Organization (WHO) from South Africa, and was soon designated as a variant of concern, subsequently named Omicron~\citep{who2021,duong2022sars}. Around the same time, the Omicron variant reached Australia. The first cases attributed to Omicron were, for example, reported by the New South Wales Health Department during the week ending 4 December, 2021 \citep{healthNSW2022}. The rapid spread of the Omicron variant dramatically changed the epidemiological situation in the country, starting a new, fourth, pandemic stage in Australia~\cite{porter2022new}. This occurred just at the time when the previous, third, wave caused by the Delta (B.1.617.2) variant, had begun to subside, having been controlled to a large extent by a comprehensive mass-vaccination campaign, as well as strict and long lockdowns across several affected states and territories (e.g., New South Wales, Victoria, Queensland, and Australian Capital Territory)~\citep{milne2022mitigating,chang2022simulating,chu2022vaccination,sanz2022modelling}. 

A traditional description of a pandemic ``wave'' assumes that a wave includes both upward and/or downward periods, during which the rising/declining case numbers are substantial~\citep{zhang2021second}. The first three pandemic waves in Australia exhibited well-defined upward and downward periods. The first wave in the nation (March--June 2020) peaked at approximately 500 cases per day, i.e., around 20 daily cases per million~\citep{chang2020modelling}. The second wave was mostly confined to the state of Victoria (June--September 2020), peaking at approximately 700 cases per day, i.e., around 30 daily cases per million~\citep{blakely2020probability,zachreson2021risk}. The third wave (June--November 2021) peaked at approximately  2,750 daily cases, i.e., around 100 daily cases per million. This peak formed by mid-October 2021, and the incidence stabilised in November between 1,200 and 1,600 daily cases, i.e., between 45 and 65 daily cases per million~\citep{covid19data,chang2022simulating}. 

However, the fourth pandemic stage caused by the Omicron variant (December 2021 -- September 2022) has included several different waves. Notably, there were multiple peaks, with the first and largest one observed in mid-January 2022 (around 110,000 daily cases, or 4,250 daily cases per million), while smaller but still substantial peaks occurred in early April 2022 (around 57,000 daily cases, or 2,200 daily cases per million) and mid-May 2022 (around 51,000 daily cases, or 2,000 daily cases per million), see Fig.~\ref{actual}. The fourth pandemic stage was also characterised by the emergence of a number of sub-lineages of the Omicron variant, some of which rapidly spread across Australia~\citep{healthNSW2022}: sub-lineage BA.1 (first reported in NSW by week 48, 2021), BA.2 (first reported in NSW by week 3, 2022), BA.4 (first reported in NSW by week 16, 2022) and BA.5 (first reported in NSW by week 17, 2022). 
Starting from March 2022,  Omicron BA.1 was mostly replaced by BA.2, and from May 2022, BA.4 and BA.5 (see Fig.~\ref{actual}).  The BA.3 sub-variant did not make a notable impact in Australia.

Different transmission rates were observed between sub-variants: BA.2 had a daily growth rate advantage relative to BA.1 of 0.10 (95\% CI: 0.10-0.11)~\cite{elliott2022twin} and BA.4 and BA.5 have an estimated  daily growth advantage of 0.08 (95\% CI: 0.08–0.09) and 0.10 (95\% CI: 0.09–0.11), respectively, relative to BA.2~\citep{tegally2022emergence}. In addition, BA.4 and BA.5 have resulted in more re-infections than BA.1 and BA.2~\citep{tuekprakhon2022antibody,NEJMc2206576,wang2022antibody}. Pandemic effects of different Omicron sub-variants  have not been traced within an individual-based  age-dependent transmission and response model that takes into account demographic heterogeneity of the Australian population. 

Age-dependent hospitalisation and fatality rates reportedly differ across all four Omicron sub-variants detected in Australia~\cite{Nyberg2022,healthNSW2022,scobie2022update}.   In addition, during the Omicron pandemic period, most hospitalisations and deaths occurred  among patients aged over 65 years, disabled patients, and those with three or more comorbidities~\cite{adjei2022mortality}.  Thus, to model the actual impact of COVID-19 on mortality and ICU admissions, cases where mortality or critical illness (ICU admission) were directly attributable to the COVID-19 must be distinguished from those where this was not the case. This differential characterisation has has not been modelled for the Omicron pandemic stage in Australia. Hence, the impact of indirect disease burden remains an open question, and providing an adequate model, validated by the actual data from the Omicron pandemic period, may elucidate future public health approaches.

The diverse epidemiological profile of the four co-existing Omicron sub-variants is complicated by fluctuations in the fraction of the Australian population that followed non-pharmaceutical interventions (NPIs), and specifically, social distancing (SD) measures. 
While during previous pandemic stages the social distancing measures in Australia were mandated by various ``stay-in-home-orders'', these restrictions were significantly relaxed during the Omicron pandemic stage. As a result, rather than considering ``SD compliance'', we focus on ``SD adoption''. Following previous studies~\cite{chang2020modelling, zachreson2021how, chang2022simulating}, we interpret social
distancing  measures as comprising several behavioural changes that reduce the intensity of interactions among individuals, including staying at home, mask wearing, physical distancing, etc, and thus, reduce the virus transmission probability (see \ref{NPIs-SD}). Consequently, the population fraction following one or more of these behaviours, that is, reducing the overall intensity of interactions, is referred to as ``SD-adopters''. 

 Substantial variability in SD-adoption during the Omicron pandemic stage is documented in several reports, including  mobility reduction reports, shown in Supplementary Fig.~\ref{fig:mobility} and Table~\ref{tab:mobility}. The mobility reduction data are categorised in a coarse-grained way (e.g., workplaces, public transport, residential, retail and recreation, etc.)~\citep{Google}. Hence, it is difficult to ascertain the extent of actual SD-adoption, i.e., the actual fraction of SD-adopters, at different times during the fourth pandemic stage, especially when interleaved with summer school holidays. It is, however, notable that during the rapid growth in incidence  towards the first peak observed in mid-January 2022, the fraction of people staying at home (not travelling to workplaces) increased. Once the incidence significantly dropped during late February 2022, this fraction was reduced. When the incidence  resumed its growth toward the second peak in mid-April, there was a slight but delayed increase in this fraction, albeit of a smaller magnitude. 

These observations of fluctuating SD-adoption indicate a degree of responsiveness within the population.  However, such responsiveness has not yet been quantified and comprehensively modelled, with previous studies assuming only static fractions of individuals compliant with (or adopting) social distancing measures. Thus, there is a need to retrospectively and quantitatively estimate the fluctuating fraction of SD-adopters, treating it as a key variable which determined the pandemic response during the Omicron stage, in presence of other NPIs and vaccination interventions.

\section{Rationale and objectives}

Overall, we aim to provide and validate a high-resolution pandemic response model on the scale of Australia. This model should be able to trace highly transmissible sub-variants of SARS-CoV-2, reproducing and explaining multiple observed waves, evaluate a range of dynamic public health interventions, and quantify their relative contributions to pandemic response. 

\subsection{Pandemic phases}

This study aims to delineate distinct phases of the fourth pandemic stage in Australia, induced by the Omicron variant. In doing so, we separate the dynamics induced by the two initial sub-variants (BA.1 and BA.2), as opposed to the subsequent sub-variants (BA.4 and BA.5).  We then model the leading part of the fourth pandemic stage, induced by the two initial sub-variants of Omicron (BA.1 and BA.2), that is, the period between early December of 2021 and mid-April of 2022 (see Fig.~\ref{actual}).   

\subsection{Fluctuating adoption of social distancing}

Independently, we also aim to identify the periods of fluctuating adoption of social distancing requirements, matching the corresponding simulations with actual pandemic curves.
In identifying the impact of the social distancing behaviour on the pandemic spread, we contrast the trajectory modelled for the fluctuating SD-adoption with its static counterparts, set at feasible extremes of SD-adoption (20\% and 70\%). 
In addition, we consider transition to the  phase dominated by sub-variants BA.4 and BA.5, and possible reasons behind persistence of the Omicron variant of SARS-CoV-2 in Australia.

\subsection{Disease burden}

Finally, we model the disease burden of the fourth pandemic stage in Australia experienced until mid-April 2022, in terms of hospitalisations, intensive care units (ICU) occupancy, and mortality caused by the Omicron variant (BA.1 and BA.2). In doing so, we account for the impacts directly attributable to, or not directly attributable to COVID-19 on these parameters. This analysis considers a balance of the fluctuating extent of SD-adoption, transitions across the sub-variants, and the indirect effects of COVID-19, and explores to what extent this balance shaped the nonlinear dynamics across the three  measures of disease burden.

\section{Methods}
\label{method}

In order to capture the heterogeneous population structure, we used a high-resolution Agent-based Model (ABM) which comprised an artificial population of over 23.4 million agents generated from the Australian census data. 
The  model was implemented within a large-scale software simulator (AMTraC-19: Agent-based Model of Transmission and Control of the COVID-19 pandemic in Australia~\citep{amtrac-code-zenodo-2022}) and included a range of non-pharmaceutical interventions and vaccination rollout schemes. This model has been previously calibrated and validated to reproduce key variant-specific epidemiological features of the COVID-19, including the ancestral variant~\cite{chang2020modelling, zachreson2021how} and the Delta variant~\cite{chang2022simulating,nguyen2022optimising}. In this work, we calibrated the model to the Omicron variant (BA.1 and BA.2), using genomic surveillance reports~\citep{healthNSW2022}; extended the social distancing model to support a fluctuating fraction of SD-adopters; carried out an additional sensitivity analysis; and validated the model using the incidence data reported between December 2021 and April 2022, as described in Supplementary Material. 

Agents are grouped in various mixing contexts which reflect the demographic features captured by the census. While contacts in ``residential'' contexts (household, household cluster, community) correspond to local interactions, other interactions occur in ``workplace/education'' mixing contexts (working groups and schools). 
An agent is initialised with a commuting pattern between a residential area and a destination zone representing workplace, using the datasets maintained by the Australian Bureau of Statistics (ABS)~\cite{zachreson2018urbanization,fair2019creating,zachreson2020interfering}. The disease transmission probabilities are set to vary across the interaction contexts and ages, see Supplementary Table~\ref{tab:transmission_probability_at_peak_infectivity}.
The transmission is simulated in discrete time, with two steps per day: ``daytime'' capturing the interactions in workplace/education contexts, and ``nighttime'' capturing  the interactions in residential contexts. 

The model implements a specific profile for the natural history of the disease, progressing over several states of infection, i.e., Susceptible, Infectious (Asymptomatic or Symptomatic), and Removed (recovered or deceased). This dynamic is governed by several parameters, including the incubation period and recovery period, which are randomised across the agents around  mean values. Susceptible hosts or agents which share a mixing context with an infectious agent may become infected and then infectious. Some agents develop asymptomatic infection, with this fraction set differently for adults and children. The asymptomatic infectivity is set lower than the symptomatic one. The model also specifies case detection probabilities which differ for symptomatic and asymptomatic/pre-symptomatic cases, see Supplementary Table~\ref{tab:main_transmission}. Sensitivity analysis is described in Supplementary Material.

A pandemic scenario is triggered by infecting a number of agents residing in metropolitan areas around international airports, in proportion to the number of incoming passengers~\cite{chang2020modelling,chang2022simulating}, as described in~\ref{ABM-prob}. 
At each simulation step, the mixing contexts of agents and their trajectory through the natural history of the disease determine their transmission probabilities. Once an agent is exposed to infection, the transmission probability varies: it exponentially grows to a peak of infectivity and then linearly declines during the agent's recovery.

In modelling the fourth pandemic stage, we assumed that the population vaccination coverage was high. This assumption is supported by the vaccination levels achieved in Australia by the end of 2021. The proportion of adult Australians (i.e., over the age of 16) who were double vaccinated by 30 November 2021 reached 87.2\%, and 92.5\% of adults have had by then at least one dose~\citep{aus--2021}. The implemented vaccination scheme follows a pre-emptive mass-vaccination campaign with two vaccines: priority and general (see Supplementary Material). This reflects key features of the actual vaccine rollout in Australia during 2021~\cite{zachreson2021how,chang2022simulating}. When an agent is vaccinated, their vaccination state is defined by a set of the corresponding vaccine efficacy parameters.

The model includes several NPIs, including case isolation, home quarantine, school closures, and social distancing behaviour~\cite{chang2020modelling,chang2022simulating}. Case isolation and home quarantine are activated from the start of simulation. The setup and duration of school closures approximates the period of summer holidays and a gradual resumption of school activities until mid-March of 2022.   
The extent of social distancing adoption is simulated dynamically, with the SD-adopting population fraction set at any given simulation step according to a predefined assignment profile, see Table~\ref{tab:dynamic_SD}. Whenever the SD-adoption fraction changes during the simulation, Bernoulli sampling determines the agents which are selected to follow the social distancing behaviour. Vaccination states, the adoption of social distancing and the compliance with other NPIs modify the transmission probabilities within the agents' mixing contexts, as described in Supplementary Material. 

Estimation of disease burden is carried out in terms of hospitalisation, ICU occupancy and mortality (daily and cumulative deaths). In doing so, we used age-dependent case hospitalisation risks (CHRs) and ICU ratios, scaled to the actual hospitalisation cases in Australia, and age-dependent infection fatality rates (IFRs), while accounting for adjusted vaccine efficacies and a differentiation between sub-variants BA.1 and BA.2, as described in Supplementary Material.

\section{Results}
\label{results}

\subsection{Pandemic phases}

\begin{figure}[!h]
    \centering
    \includegraphics[width=0.9\columnwidth,trim=1cm 4cm 1cm 4cm,clip]{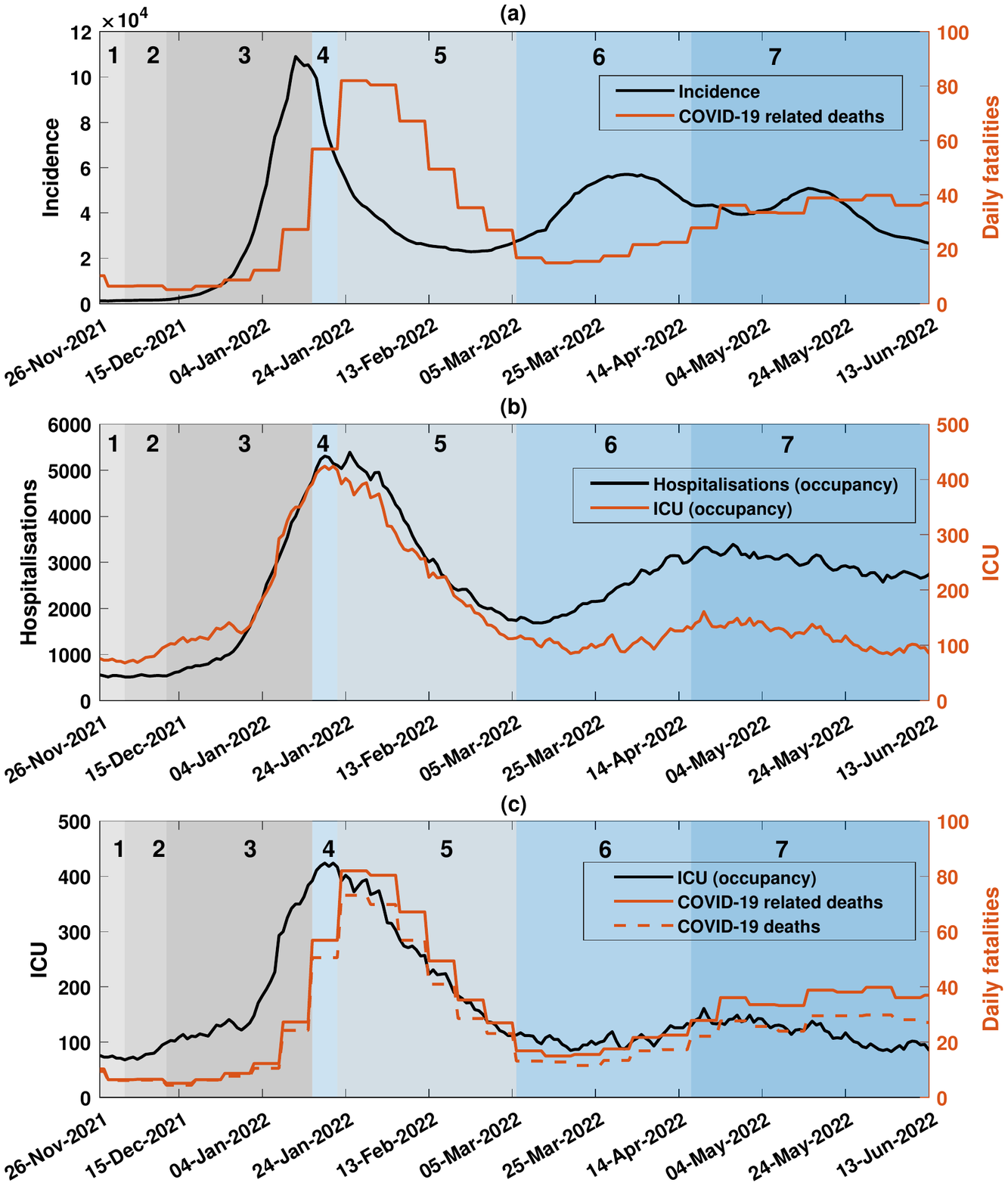}
    \caption{Incidence and disease burden of COVID-19 in Australia between 26th November 2021 and 13th June 2022. Shaded areas in grey and blue show the emergence of variants of concern and sub-lineages over time, identified in weekly genomic surveillance reports (NSW Health). Top (a): Black (y-axis, left): a 7-day moving average of reported COVID-19 daily incidence. Orange (y-axis, right): COVID-19 related deaths. Middle (b): Black (y-axis, left): COVID-19 hospitalisations (bed occupancy). Orange (y-axis, right): COVID-19 ICU cases (bed occupancy). Bottom (c): Black (y-axis, left): COVID-19 ICU cases (bed occupancy); Orange (y-axis, right): Daily COVID-19 related deaths (solid), daily COVID-19 deaths (dashed).  The data for COVID-19 cases, hospitalisation and ICU occupancy, and mortality are published by the Australian government. }
    \label{actual}
\end{figure}

Using contemporary genomic surveillance reports~\citep{healthNSW2022}, we identified seven main phases of the fourth pandemic stage in Australia caused by the Omicron variant, shown in Figures~\ref{actual} and~\ref{incidence}. The timeline is divided into 7 phases as follows:
\begin{itemize}
    \item Phase 1: BA.1 detected, epidemiological week 48, ending 4 December, 2021.
    \item Phase 2: Delta and BA.1 co-exist, epidemiological week 49, ending 11 December, 2021. 
    \item Phase 3: BA.1 dominant, epidemiological week 50 (2021), ending 18 December, 2021, to week 2 (2022).  
    \item Phase 4: BA.2 detected, epidemiological week 3, ending 22 January, 2022.
    \item Phase 5: BA.1 and BA.2 co-exist, epidemiological week 4 to week 9, ending 5 March, 2022.
    \item Phase 6: BA.2 dominant, epidemiological week 10, ending 12 March, 2022.
    \item Phase 7: BA.2, BA.4 and BA.5 co-exist, epidemiological week 16, ending 23 April, 2022, onwards\footnote{Data source: COVID-19 weekly surveillance reports, published by NSW Health~\citep{healthNSW2022}.}.
\end{itemize}
We note that sub-variant BA.1 has displaced the Delta variant by phase 3,  causing a rapid growth in incidence towards the primary incidence peak. The second incidence peak shaped during phase 6 which was dominated by sub-variant BA.2. Phase 7 marked  the emergence of BA.4 and BA.5.

We re-calibrated the ABM to the incidence data observed during the beginning of the fourth pandemic stage in Australia, driven by the Omicron variant (specifically, phase 3 in Fig.~\ref{actual} shaped by BA.1, but considering also phases 4 to 6 when BA.1 was concurrent with BA.2). For the previous pandemic stage in Australia (June--November 2021), i.e., the third pandemic wave shaped by the Delta variant, the reproductive number was calibrated to be around $R_0 = 6.20$~\cite{chang2022simulating}. In this study, the calibration resulted in $R_0 = 19.56$ (see Supplementary Material, including \ref{calibr-sens} and Table~\ref{tab:epi_parameters}), which is consistent with other reports estimating $R_0$ of Omicron BA.1 to be approximately 3.1 times as high as the reproductive number of the Delta variant~\cite{obermeyer2022analysis}.

\subsection{Fluctuating adoption of social distancing}

In modelling the first six phases during which the two initial sub-variants of Omicron (BA.1 and BA.2) emerged and reached dominance, we focused on a ``retrodictive'' estimation of the  SD-adopting population fraction. A key assumption that we made is that the adoption of social distancing behaviour fluctuates over time and the changes in the extent of SD-adoption are not necessarily synchronous with the pandemic phases.  Instead, these changes can be driven by individual perception of the infection risks, life pattern changes associated with summer holidays and return to school/work for many individuals, vaccination availability for children, and other policy setting changes affecting social distancing. Hence the changes in the extent of SD-adoption may precede or follow some salient observed features of the pandemic trajectory. 

\begin{figure}[!h]
    \centering
    \includegraphics[width=\columnwidth,trim=2cm 1cm 3cm 1cm,clip]{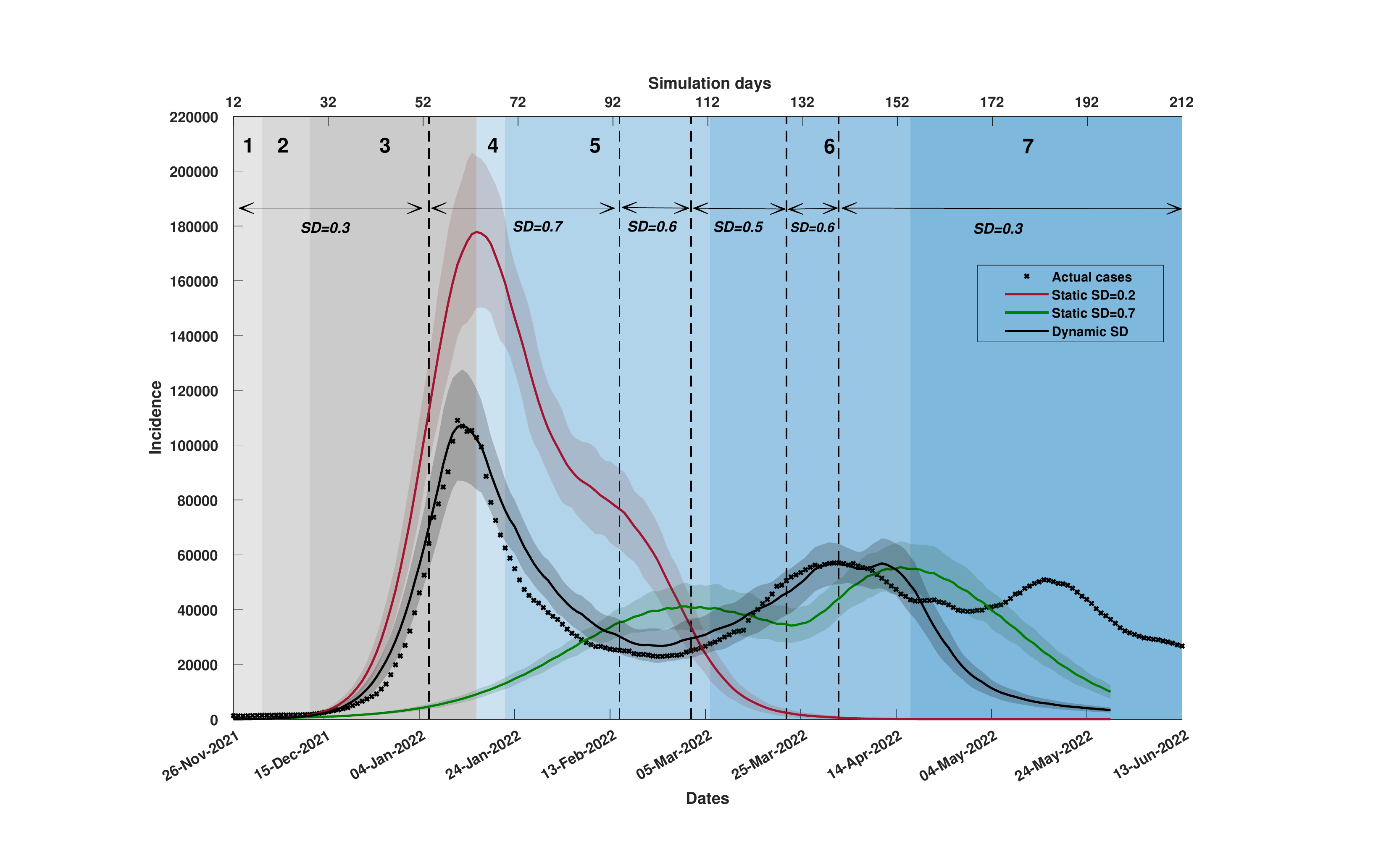}
    \caption{A comparison of incidence produced by dynamic social distancing (SD) adoption (shown in solid black) and static SD-adoption fractions ($SD_1 = 0.2$, shown in red; $SD_2 = 0.7$, shown in green). Coloured shaded areas around the solid line show standard deviation. Changes in dynamic SD-adoption are marked by vertical dashed black lines. Traces corresponding to each simulated scenario are computed as the average over 20 runs. Simulated SD adoption is combined with other interventions (i.e., school closures, case isolation, and home quarantine). A 7-day moving average of the actual time series (black crosses) is shown for the period between 26th November 2021 and 13th June 2022 (x-axis, bottom). The simulated incidence is scaled up by 10\% to reflect the population increase from 23.4 million (2016 census data, model input) to 25.8 million (2021 census data). The simulated incidence is offset by 12 days (x-axis, top) to align with the observed incidence peak. 
    Shaded areas in grey and blue show the emergence of variants of concern and sub-lineages over time, identified in weekly genomic surveillance reports (NSW Health).}
    \label{incidence}
\end{figure}

Selection of the SD-adoption change-points and the corresponding fractions of SD-adopters is a subject of modelling which aims to minimise the difference between the resulting and observed incidence curves. For example, the SD-adoption profile presented in Table~\ref{tab:dynamic_SD} includes six change-points, including the initial assignment, at day 0, of $SD = 0.3$, that is, 30\% of the agents follow social distancing behavior from the pandemic stage onset. This is set to continue until day 54 after which the fraction of SD-adopters sharply rises to 70\%, i.e., $SD = 0.7$ from day 55, and so on. This optimised profile produced the average incidence curve shown in Fig.~\ref{incidence}, closely approximating the observed incidence curve, with both primary and secondary peaks well-aligned with the observations. The optimisation process is described in Supplementary Material.

The incidence curve produced by the fluctuating (dynamic) SD-adoption is contrasted with two static SD-adoption alternatives, set at 20\% and 70\%, i.e., $SD_1 = 0.2$ and $SD_2 = 0.7$. 
This comparative analysis, illustrated by Fig.~\ref{incidence}, highlights the nonlinear nature of  resultant dynamics: the fluctuating fraction of SD-adopters produces the trajectory which agrees with the actual incidence dynamics, unlike the static SD-adoption profiles which both fail to match the salient pandemic patterns. For example, the lower static SD-adoption, $SD_1 = 0.2$, does not produce the secondary peak at all, and the higher static SD-adoption, $SD_2 = 0.7$, misses the first peak by almost two months and produces the second peak with a larger magnitude than the first one (see Supplementary Table~\ref{tab:inc_peak}). 

There was a notable divergence between the simulated and observed curves, which developed  at the onset of phase 7, in early May 2022 (Fig.~\ref{actual}). The incidence observed in phase 7 correlated with the spread of sub-variants BA.4 and BA.5, and the daily cases reached substantial levels, generating a third peak at the end of May. In contrast, the simulated curve fell during this period to near-zero levels. 
We note that no extent of the SD-adoption fraction simulated between $SD_{min} = 0$ and $SD_{max} = 0.7$, was sufficient to maintain the infection spread.  This can be explained by depletion of the susceptible host population within the simulation: the simulated cumulative incidence at the end of phase 6 reached 23\% of the simulated population, with the rest being probabilistically protected by vaccinations (given  the partial efficacy of both vaccines). Since the model does not assume re-infections and does not simulate diminishing vaccine efficacy, this divergence strongly suggests that the actual incidence observed during phase 7, associated with the spread of BA.4 and BA.5 (including the third peak, see Fig.~\ref{actual}), was mostly generated by re-infections or new infections in vaccinated individuals with lower vaccine effectiveness (e.g., due to waning vaccine efficacy), rather than infections of the epidemiologically  or immunologically ``naïve'' population (see Discussion).

\subsection{Disease burden}

We next modelled the disease burden generated  by the Omicron variant (BA.1 and BA.2), measured in terms of (i) hospitalisations (occupancy), (ii)  ICU (occupancy), and (iii) mortality (daily and cumulative deaths) in Australia, as described in Supplementary Material. The key input to the disease burden estimation is provided by the incidence curves produced by the simulation. In other words, a given SD-adoption profile generates a non-linear incidence trajectory which in turn results in hospitalisations and ICU occupancy. The incidence trajectory also leads to estimation of daily and cumulative deaths. Similarly to the analysis of daily incidence, we contrast dynamic and static SD adoption. The difference, however, is that we no longer optimise the SD assignment profile, and use the optimised profile presented in Table~\ref{tab:dynamic_SD}. Thus, a comparison between the actual and estimated disease burdens can be used to further validate the model. 

\begin{figure}
    \centering
    \includegraphics[width=\columnwidth,trim=2cm 1cm 3cm 1cm,clip]{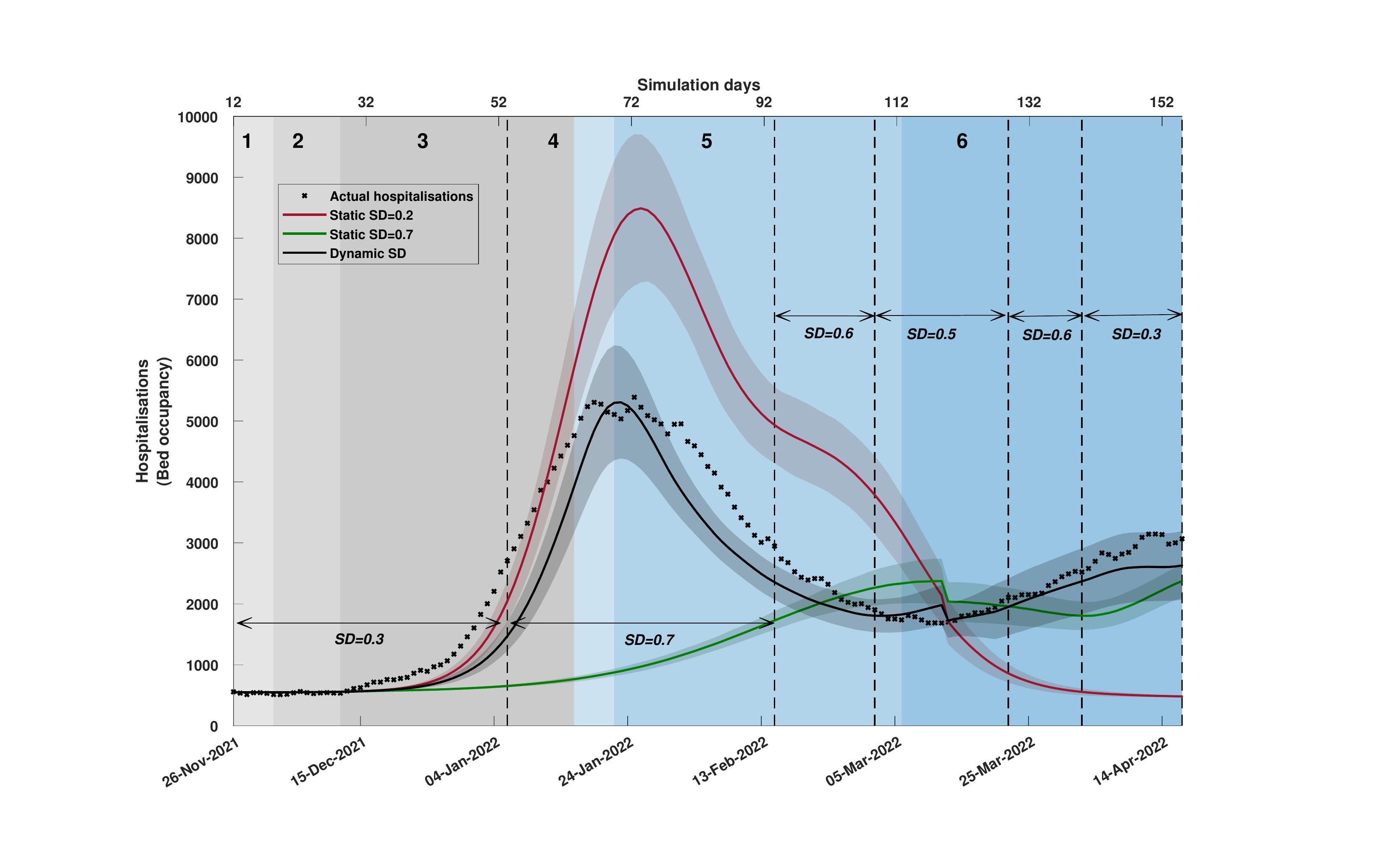}
    \caption{A comparison of hospitalisations (bed occupancy) among dynamic social distancing (SD) levels (shown in solid black) 
    and static SD levels ($SD_1 = 0.2$, shown in red; $SD_2 = 0.7$, shown in green). The simulated hospitalisations are offset by 7 days. Coloured shaded areas around the solid line show standard deviation. Changes in dynamic SD-adoption are marked by vertical dashed black lines. Traces corresponding to each simulated scenario are computed as the average over 20 runs.  SD adoption is combined with other interventions (i.e., school closures, case isolation, and home quarantine). The actual time series (black crosses), shown from 26th November 2021, aligns with the start of the Omicron outbreak in Australia. Shaded areas in grey and blue show the emergence of variants of concern and sub-lineages over time, identified in weekly genomic surveillance reports (NSW Health). }
    \label{hosp}
\end{figure}

Figure~\ref{hosp} and Supplementary Table~\ref{tab:clinical_peak} show that the hospitalisation curve (occupancy) generated by the dynamic SD-adoption closely matches the actual hospitalisations. 
This alignment is again in stark contrast with the hospitalisation curves produced by static SD-adoption fractions ($SD_1 = 0.2$ and $SD_2 = 0.7$).

\begin{figure}
    \centering
    \includegraphics[width=\columnwidth,trim=2cm 1cm 3cm 1cm,clip]{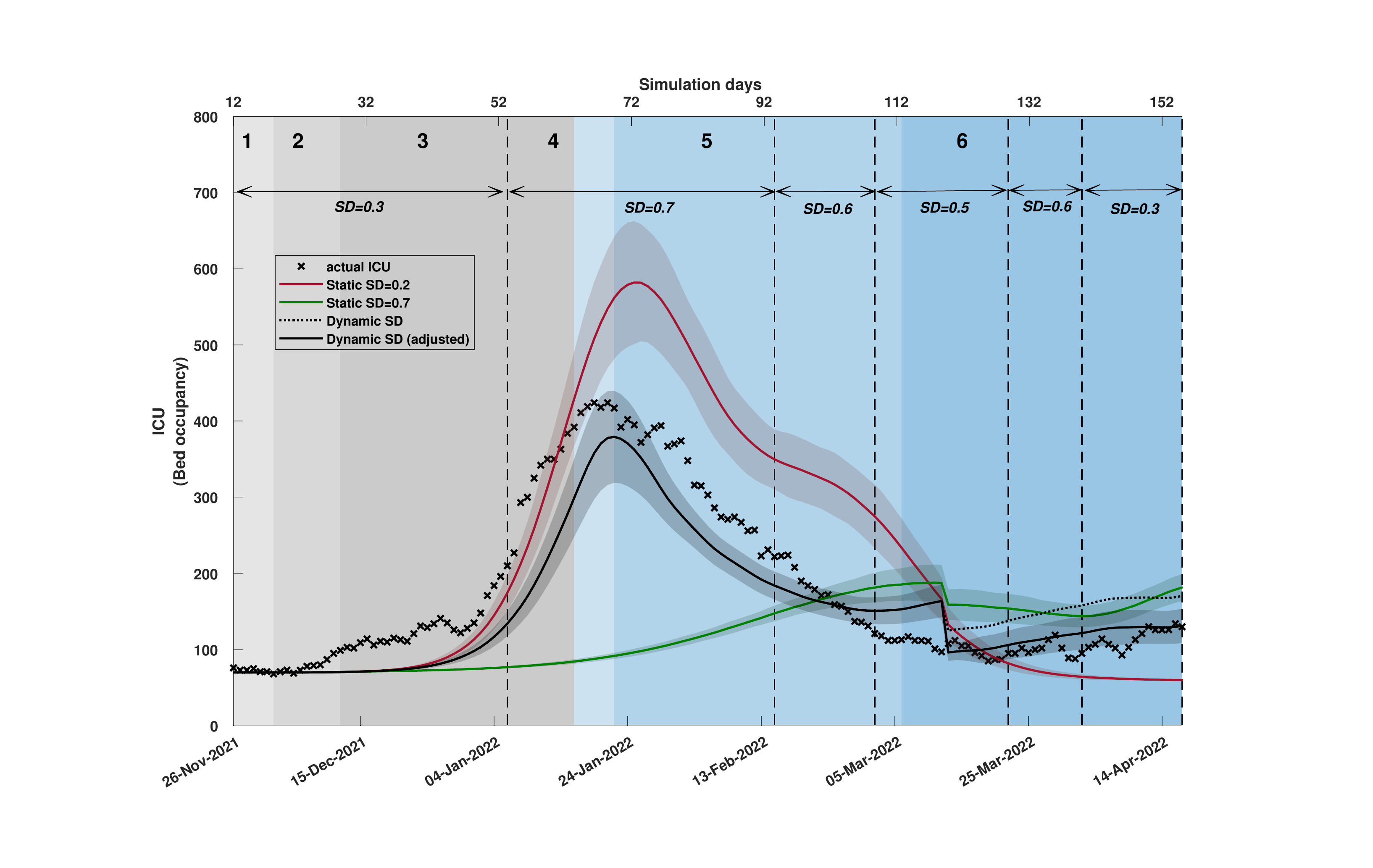}
    \caption{A comparison of ICU (bed occupancy) among (i) dynamic social distancing (SD) levels, adjusted for phase 6 by scaling with $0.77$ to capture ICU cases due to COVID-19 (shown in solid black), (ii) non-adjusted dynamic SD levels to represent all ICU cases (dotted black), and (iii) static SD levels ($SD_1 = 0.2$, shown in red; $SD_2 = 0.7$, shown in green). The simulated ICUs are offset by 7 days. Coloured shaded areas around the solid line show standard deviation. Changes in dynamic SD-adoption are marked by vertical dashed black lines. Traces corresponding to each simulated scenario are computed as the average over 20 runs.  SD adoption is combined with other interventions (i.e., school closures, case isolation, and home quarantine). The actual time series (black crosses), shown from 26th November 2021, aligns with the start of the Omicron outbreak in Australia. Shaded areas in grey and blue show the emergence of variants of concern and sub-lineages over time, identified in weekly genomic surveillance reports (NSW Health).}
    \label{ICU}
\end{figure}

Figure~\ref{ICU} and Supplementary Table~\ref{tab:clinical_peak}  show that the ICU curve (occupancy) generated by the dynamic SD-adoption  matches the actual ICU occupancy reasonably well during the initial five phases which were shaped mostly by the BA.1 sub-variant, and somewhat diverges during phase 6 dominated by BA.2.  During phase 6 (that is, for the last six weeks of the considered timeline), the simulated ICU occupancy is higher than the actual ICU occupancy. 
The ICU occupancy model, driven by  ICU admission rates, accounted for a  reduction in the disease severity,  differentiating between BA.1 and BA.2 (cf. Supplementary Table~\ref{tab:vac_scalar}). Furthermore, we adjusted the ICU trajectory during phase 6 to represent the ICU occupancy directly attributable to COVID-19, rather than the ICU occupancy in patients with COVID-19 (as explained in \ref{mort}). The adjusted ICU occupancy is in a better concordance with the actual data.
The static SD-adoption fractions ($SD_1 = 0.2$ and $SD_2 = 0.7$) do not capture the actual trajectory of ICU occupancy.

\begin{figure}
    \centering
    \includegraphics[width=\columnwidth,trim=2cm 1cm 3cm 1cm,clip]{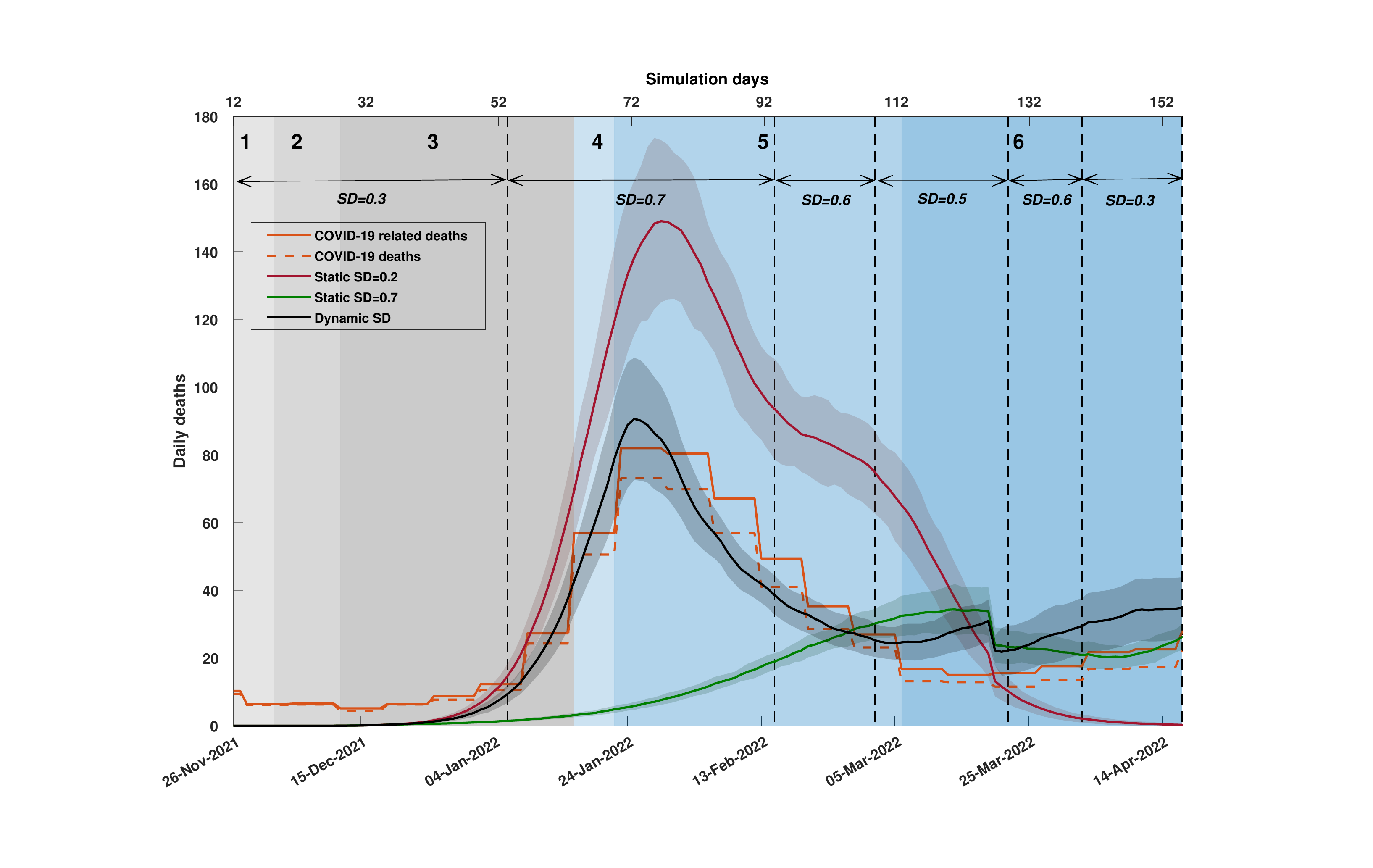}
    \caption{A comparison of daily deaths between dynamic social distancing (SD) levels (shown in solid black) and static SD levels ($SD_1 = 0.2$, shown in red; $SD_2 = 0.7$, shown in green). The simulated daily deaths are offset by 14 days. Coloured shaded areas around the solid line show standard deviation. Changes in dynamic SD-adoption are marked by vertical dashed black lines. Traces corresponding to each simulated scenario are computed as the average over 20 runs. SD adoption is combined with other interventions (i.e., school closures, case isolation, and home quarantine). Actual daily deaths are derived from reported weekly deaths (shown in orange; solid: COVID-19 related deaths; dashed: COVID-19 deaths). Shaded areas in grey and blue show the emergence of  variants of concern and sub-lineages over time, identified in weekly genomic surveillance reports (NSW Health). The timeline is divided into 6 phases as follows: 1) BA.1 detected, 2) Delta and BA.1 co-exist, 3) BA.1 dominant, 4) BA.2 detected, 5) BA.1 and BA.2 co-exist, and 6) BA.2 dominant.}
    \label{daily_death}
\end{figure}

Finally, Fig.~\ref{daily_death}, Fig. ~\ref{cum_death} and Supplementary Table~\ref{tab:clinical_peak} show that the daily mortality curve produced by the dynamic SD-adoption  matches the actual data reasonably well again. In contrast,  the mortality estimated by static SD-adoption fractions, $SD_1 = 0.2$ and $SD_2 = 0.7$, consistently overestimate and underestimate the actual observations, respectively. 

These estimations of the disease burden reinforce the argument that the nonlinear dynamics of the observed hospitalisations, ICU occupancy, and mortality result from a nuanced combination of (i)  fluctuating fractions of SD-adoption,  (ii) a transition from BA.1 to BA.2,  and (iii)  indirect effects of COVID-19, i.e., the distinction between  ICU admissions or deaths arising due to COVID-19 versus those with COVID-19.

\section{Discussion}
\label{discussion}

Emergence of recurrent epidemic waves of COVID-19 is a systemic phenomenon and its quantitative understanding remains an elusive target~\cite{campi2022sars}. Typically, several interacting factors contribute to recurrent dynamics.
Firstly, the population structure in the modern world is highly heterogeneous: not only at the level of individual households and local government areas, but also at the regional, national and international level. This heterogeneity forces the infection transmission to follow complex paths, producing distinct urban and rural waves~\citep{aneja2021assessment,zachreson2018urbanization,selvaraj2022rural}, disproportionate effects on high- and low-density housing~\citep{rasul2021socio}, as well as in the areas characterised by socioeconomic disadvantage profiles and higher concentrations of essential workers~\citep{gilman2020modelling,chang2022simulating,espana2022designing,aylett2022epidemiological}. Thus, while infection rates may decline in some areas, they may escalate in others. 

Secondly, the population compliance with various stay-at-home orders and adoption of social distancing requirements fluctuates in time: strong initial adherence is typically replaced by sizable fatigue and rarely recovers beyond moderate compliance, thus reducing effectiveness of NPIs~\citep{pedro2020conditions,chang2022simulating,delussu2022evidence}. This fatigue cannot be exclusively attributed to ``contrarian'' individuals and affects broad population groups. 

Thirdly,  vaccine efficacy wanes over time, requiring booster vaccine doses, which temporarily reestablish protective immunity. Independently, following a typically strong initial vaccination uptake,  ``the fully up to date''  vaccinated population fraction slowly reduces over time, undergoing possible upswings when the risk perception increases~\citep{bauch2004vaccination,bauch2012evolutionary}. Similar to the NPI fatigue, this dynamic is driven not only by ``anti-vaxxer'' groups, but may depend on access to and knowledge about vaccination~\cite{attwell2022covid}, and involve several drivers of vaccine hesitancy, including confidence, complacency, altered risk calculation, and limited collective responsibility~\cite{machingaidze2021understanding,chang2019effects}.

In addition, the testing, tracing, isolation and quarantine (TTIQ) capacities themselves may be under stress during rapidly growing outbreaks, diminishing their role in curbing the epidemic. The TTIQ capacities were severely strained in Australia during the rise of the Omicron variant, leading to a decline in their effectiveness~\citep{AHPPC2021}. 
Moreover, delays in imposing strict NPIs or initiating vaccination rollouts amplify non-linearly, so that short response lags cause long recovery tails~\citep{chang2020modelling,chang2022simulating}. All these cyclical spatiotemporal dynamics may combine over time, creating vicious feedback loops and superposition of interacting waves. 

In this study, we sought to increase our understanding of the mechanisms generating multiple pandemic waves. We specifically modelled a pandemic response to the emergence and spread of highly transmissible variant of SARS-CoV-2 (Omicron and its sub-lineages) during the fourth pandemic stage in Australia. This stage produced several local incidence peaks, and we identified the corresponding phases (see Fig.~\ref{actual}), followed by retrospective modelling of dynamic social distancing behaviour. 

Importantly, the identification of distinct phases did not reduce the analysis to a description of well-defined waves, but rather attributed the incidence peaks and patterns to nuanced combinations of new sub-variants and fluctuations in social distancing behaviour. For example, the secondary peak in mid-April of 2022 (see phase 6 in Fig.~\ref{actual}) was explained not only by the emergence of BA.2 which started to dominate the preceding BA.1, but also by the reduced adoption of social distancing requirements.

The main result of this modelling  is that the fraction of SD-adopters was found to be fluctuating in response to the incidence dynamics. Essentially, these changes appeared in response to the  pandemic dynamics observed by the population, sometimes lagging behind and occasionally preempting pandemic turns. 

Using the actual incidence data and simulated scenarios which varied the extent of SD-adoption across time, we ``retrodictively'' determined the dynamically fluctuating SD-adoption profile that produced the incidence curves  matching the observations. The resultant incidence curves were further contrasted with the curves produced by alternative scenarios defined by static SD-adoption. The comparison clearly demonstrated that, unlike the dynamic SD-adoption, the static alternatives failed to reproduce key non-linear features of the incidence trajectories. This indicated that the fraction of individuals adopting (or complying with) social distancing requirements may greatly vary over time, especially during a long pandemic. Hence, public health policies must adequately account for this variability, acknowledging and addressing possible fatigue and complacency within the population. Timely, rapidly designed and focused communication campaigns may be needed to reinforce the importance of continuing social distancing in controlling persistent spread and reducing the public health burden.  

The study also suggested, albeit indirectly, that the increased incidence observed during the last wave induced by sub-variants BA.4 and BA.5 (Fig.~\ref{actual}: phase 7) was mostly driven by re-infections or new infections in vaccinated individuals with low or waning vaccine effectiveness, rather than new infections of the epidemiologically or immunologically ``naïve'' population. Testing for anti-spike antibodies in Australia  between 9 and 18 June 2022 provided an indication of the extent of cumulative exposure in the community to vaccination and/or natural infection: the prevalence of anti-spike antibodies, elicited by both vaccination and natural infection, was found to be very high (99\%) across all jurisdictions, while the prevalence of anti-nucleocapsid antibodies, elicited by natural infection, reached 46\%~\cite{lifeblood2022seroprevalence}. The finding of our study, which emphasised re-infections and low or diminishing vaccine efficacy as key factors during phase 7, pointed to the need for agile ``booster'' vaccination campaigns, to raise immunity levels within the population during protracted pandemic stages. 

We estimated the corresponding disease burden (the hospitalisations, ICU occupancy, and daily and cumulative deaths), and compared it with the actual data. A good agreement between the simulated and actual disease burden data  reinforced the point that the observed nonlinear dynamics are produced not simply by transitions across sub-variants of Omicron, but also in response to varying adoption of social distancing behaviour. The analysis also quantified and emphasised indirect effects of COVID-19 on the disease burden in Australia, differentiating between ICU cases  directly attributable to COVID-19 versus those occurring with COVID-19. This distinction highlighted the need for a precise attribution of underlying conditions associated with ICU admissions during a long pandemic.

 Several limitations of the study have to be acknowledged. Firstly, the simulated population was generated using the latest available Australian Census
data from 2016 (23.4M individuals) and  may not be precisely representative of 2021--2022 period of time. However, the generated curves were scaled by 10\% to reflect the current Australian population (25.8M). 
Secondly, the ABM does not simulate  transmissions  within healthcare facilities. However,  health care professionals were vaccinated over several priority phases carried out in Australia in 2021 and 2022 and mask mandates were in place in health care facilities in these time periods, reducing the effect of this limitation. 
Furthermore, while vaccine efficacy is known to diminish over time, we did not model this explicitly, and considered this effect to be balanced by third vaccination doses administered in Australia during the studied period (e.g., 71.6\% of Australian adults had received 3 doses by August 2022). In contrast to the greatly fluctuating SD-adoption, a decline in the vaccine efficacy is known to be monotonic~\cite{szanyi2022log}, while our focus was on modelling multiple peaks of distinct waves within a single pandemic stage.
 
In conclusion, the understanding of the impact of the Omicron variant on the COVID-19 pandemic continues to be refined, and our results may benefit from  additional analysis of the latest Omicron sub-variants. Nevertheless, the quantitative retrospective modelling presented in this study highlighted the impact of fluctuating social distancing behaviour on the resultant pandemic dynamics, revealing  nuanced reasons for persistence of the Omicron variant in Australia.  
The developed model, validated with actual data from the Omicron stage in Australia, captures multiple dynamic factors which non-linearly affect the pandemic spread, and hence, can be used to evaluate and contribute to agile, adaptive and multi-faceted public health responses in the future.

\section*{Declaration of interest}
The authors declare that they have no competing interests.

\section*{Author Contributions}

 SLC, QDN and MP designed  computational experiments and re-calibrated the agent-based model. QDN and MP designed and simulated dynamic social distancing.  SLC and MP modelled disease burden. SLC and QDN carried out  computational experiments and verified the underlying data. SLC carried out the sensitivity analysis and prepared all figures. SLC, QDN and MP had full access to the data in the study. MP supervised the study. All authors contributed to drafting and editing of the Article and read and approved the final Article.

\section*{Funding}
This work was  supported by the Australian Research Council grant DP220101688 (MP, VS, TCS, SLC and QDN). AM is salary funded by an NHMRC Investigator Grant. 
AMTraC-19 is registered under The University of Sydney’s invention disclosure CDIP Ref. 2020-018. 

\section*{Acknowledgments}
We are thankful for support provided by High-Performance Computing (HPC) service (Artemis) at the University of Sydney.

\section*{Data Availability Statement}
We used anonymised data from the 2016 Australian Census obtained from the Australian Bureau of Statistics (ABS) and the Australian
Curriculum and Assessment and Reporting Authority (ACARA). These datasets can be obtained publicly, with the exception of the work travel data which can be obtained from the ABS on request. It should be noted that some of the data needs to be processed using the TableBuilder: https://www.abs.gov.au/websitedbs/censushome.nsf/home/tablebuilder.  The actual incidence data are available from the health departments across Australia (state, territories, and national), and at: https://www.covid19data.com.au/. Other source and supplementary data are available at Zenodo~\citep{amtrac-data-zenodo-2022}. The source code of AMTraC-19 is also available at Zenodo~\citep{amtrac-code-zenodo-2022}.

\appendix
\section{Supplementary material}

\subsection{Agent-based model and transmission probabilities}
\label{ABM-prob}

In simulating the transmission and control of the COVID-19 pandemic on the scale of Australia, we follow the agent-based modeling (ABM) approach, with more than 23.4M agents generated using data from Australian Census and other data provided by the Australian Bureau of Statistics (ABS) and the Australian Curriculum and Assessment and Reporting Authority (ACARA)~\cite{amtrac-code-zenodo-2022,acemod-code-zenodo-2021}.

During a ``seeding'' phase, new daily infections are generated using a binomial probability distribution and the international air traffic data from the Australian Bureau of Infrastructure, Transport, and Regional Economics (BITRE). These infections are then assigned to agents residing in randomly selected areas within a certain radius from an airport~\cite{cliff2018nvestigating}.  For example, newly generated daily infections within Greater Sydney's boundaries are distributed within a 50 km radius of Sydney's international airport.

Agent states include: Susceptible, Latent, Infectious (asymptomatic or symptomatic), and Removed (recovered or deceased). The set $G_i$ contains all mixing contexts (e.g., residential, workplace, etc.) of agent $i$.  At time step $n$, the infection probability for susceptible agent $i$ across context $g \in G_i$ is determined as follows:
\begin{equation}
    p^g_i(n) = 1 - \prod_{j \in A_g\backslash\{i\}} (1 - p^g_{j \rightarrow i}(n))
    \label{eq:infection_prob_within_a_group}
\end{equation}
where $A_g\backslash\{i\}$ is the list of agents in the context $g \in G_i$ excluding agent $i$, and  $p^g_{j \rightarrow i}(n)$ is the instantaneous probability that an infectious agent $j$, sharing the context $g$ with susceptible agent $i$, transmits the infection to agent $i$:
\begin{equation}
\label{eq:individual_transmission_prob}
    p^g_{j \rightarrow i}(n) = \kappa \ f(n-n_j | j) \ q^g_{j \rightarrow i}    
\end{equation}
A global transmission scalar $\kappa$ calibrates the reproductive number $R_0$ (see sensitivity analysis in section \ref{sens-kappa}.). The step $n_j$ marks the time when agent $j$ becomes infected, and a function $f(n-n_j| j)$ represents the natural history of the disease, i.e., the infectivity of agent $j$ over time. For an uninfected agent $j$, $n < n_j$ and $f(n-n_j | j)=0$. For an infected agent $j$, $n  \geq n_j$ and $f(n-n_j| j) \geq 0$. The infectivity increases exponentially  until its peak, $f(n-n_j | j) = 1.0$. After the peak, during the recovery period, the infectivity   decreases linearly to 0.0, when the agent state changes to Removed. The age- and context-dependent daily probabilities of transmission from agent $j$ to agent $i$, denoted by $q^g_{j \rightarrow i}$, are specified in Table~\ref{tab:transmission_probability_at_peak_infectivity}, following prior studies~\cite{chang2022simulating,nguyen2022optimising}. 

Asymptomatic agents are  modeled to be less infectious than symptomatic agents, with their infectivity scaled down by factor  $\alpha_{asymp}$. The sensitivity of the model to changes in this factor is explored in section \ref{sens-alpha}. 

\begin{table}[h]
    \centering
    \begin{tabular}{l|l|l}
    Mixing context &  Type of interaction & Daily transmission probability ($q^g_{j \rightarrow i}$)\\
        \hline
    \hline
    \rule{0pt}{3ex}Household (size 2)   & Any to child (0 - 18)             &  \hspace{1.7cm} 0.09335\\
                                        & Any to adult (19+)                &  \hspace{1.7cm} 0.02420\\
    \rule{0pt}{3ex}Household (size 3)   & Any to child (0 - 18)             &  \hspace{1.7cm} 0.05847\\
                                        & Any to adult (19+)                &  \hspace{1.7cm} 0.01495\\
    \rule{0pt}{3ex}Household (size 4)   & Any to child (0 - 18)             &  \hspace{1.7cm} 0.04176\\
                                        & Any to adult (19+)                &  \hspace{1.7cm} 0.01061\\
    \rule{0pt}{3ex}Household (size 5)   & Any to child (0 - 18)             &  \hspace{1.7cm} 0.03211\\
                                        & Any to adult (19+)                &  \hspace{1.7cm} 0.00813\\
    \rule{0pt}{3ex}Household (size 6)   & Any to child (0 - 18)             &  \hspace{1.7cm} 0.02588\\
                                        & Any to adult (19+)                &  \hspace{1.7cm} 0.00653\\[0.1cm]
    
    \hline                              
    \rule{0pt}{3ex}Household Cluster    & Child (0 - 18) to child (0 - 18)  &  \hspace{1.7cm} 0.00400\\
                                        & Child (0 - 18) to adult (19+)     &  \hspace{1.7cm} 0.00400\\
                                        & Adult (19+)  to child (0 - 18)    &  \hspace{1.7cm} 0.00400\\
                                        & Adult (19+)  to adult (19+)       &  \hspace{1.7cm} 0.00400\\[0.1cm]
    \hline
    \rule{0pt}{3ex}Working Group        & Adult (19+)  to adult (19+)       &  \hspace{1.7cm} 0.00400\\[0.1cm]
    
    \hline
    \rule{0pt}{3ex}School               & Child (0 - 18) to child (0 - 18)  &  \hspace{1.7cm} 0.00029\\
    Grade                               & Child (0 - 18) to child (0 - 18)  &  \hspace{1.7cm} 0.00158\\
    Class                               & Child (0 - 18) to child (0 - 18)  &  \hspace{1.7cm} 0.00865\\[0.1cm]
                           
    \hline
    \rule{0pt}{3ex}Neighborhood         & Any to child (0 - 4)          &  \hspace{1.7cm} $0.035 \times 10^{-5}$\\
                                        & Any to child (5 - 18)         &  \hspace{1.7cm} $1.044 \times 10^{-5}$\\
                                        & Any to adult (19 - 64)        &  \hspace{1.7cm} $2.784 \times 10^{-5}$\\
                                        & Any to adult (65+)            &  \hspace{1.7cm} $5.568 \times 10^{-5}$\\[0.1cm]
    
    \hline
    \rule{0pt}{3ex}Community            & Any to child (0 - 4)          &  \hspace{1.7cm} $0.872 \times 10^{-6}$\\
                                        & Any to child (5 - 18)         &  \hspace{1.7cm} $2.608 \times 10^{-6}$\\
                                        & Any to adult (19 - 64)        &  \hspace{1.7cm} $6.960 \times 10^{-6}$\\
                                        & Any to adult (65+)            &  \hspace{1.7cm} $13.92 \times 10^{-6}$\\[0.1cm]
    
    \hline
    \end{tabular}
    \caption{Daily transmission probabilities $q^g_{j \rightarrow i}$ from infected agent $j$ to susceptible agent $i$ for different mixing contexts and interaction types. Numbers in brackets show age groups.}
    \label{tab:transmission_probability_at_peak_infectivity}
\end{table}

The infection probability for agent $i$  across all mixing contexts is determined as follows:
\begin{equation}
\begin{split}
    p_i(n)  &= 1 - \prod_{g \in G_i(n)} \left( 1 - p^g_i(n) \right) \\
            &= 1 - \prod_{g \in G_i(n)} \prod_{j \in A_g\backslash\{i\}} \left( 1 - p^g_{j \rightarrow i}(n) \right)
\end{split}
\label{eq:general_infection_prob}
\end{equation}
At the end of each time step, the probability $p_i(n)$ is used by Bernoulli sampling to determine whether a susceptible agent $i$ becomes infected.

The probability of symptomatic illness is then calculated by adjusting the infection probability $p_i(n)$  by  a scaling factor $\sigma$ representing the fraction of symptomatic cases over the total cases:
\begin{equation}
    p_i^d(n) = \sigma(i) \ p_i(n)
    \label{eq:infection_prob_symptomatic}
\end{equation}
The fraction $\sigma(i)$ is specified for adults ($\text{age} \geq 18$), $\sigma_a = 0.67$, and children ($\text{age} < 18$), $\sigma_c = 0.268$, following prior studies~\cite{chang2022simulating}, and is varied in sensitivity analysis, see section \ref{sens-sigma}. The main parameters of the ABM are summarised in Tables~\ref{tab:main_transmission} and~\ref{tab:epi_parameters}.

\subsection{Non-pharmaceutical Interventions}
\label{NPIs-SD}
The  model includes several non-pharmaceutical interventions (NPIs): case isolation (CI), home quarantine (HQ), school closures (SC), and social distancing (SD). Each NPI is defined by (i) the population fraction that adopts it, and (ii) the adjusted (typically, decreased) strengths of interactions between an NPI-adopting agent and other agents within their mixing groups, see Table~\ref{tab:NPI}. The infection probability $p_i(n)$ for NPI-adopting agents is adjusted as follows:
\begin{equation}
    p_i(n) =  1 - \prod_{g \in G_i(n)} \left [1 - F_g(i) \left ( 1 - \prod_{j \in A_g\backslash\{i\}} (1 - F_g(j) \ p^g_{j \rightarrow i}(n)) \right ) \right ]
    \label{eq:infection_prob_compliant_agent}
\end{equation}
where $F_g(j) \neq 1$ is the strength of the interaction between agent $j$ and other agents in the mixing context $g$. For non-adopting agents $j$, the interaction strength is unchanged: $F_g(j) = 1$.  

For each agent $j$ adopting multiple NPIs, the value of $F_g(j)$ is preferentially assigned to only one NPI in accordance with the following order: CI, HQ, SD, SC. At each time step and for each NPI, the NPI-adopting and non-adopting agents  are randomly selected  according to Bernoulli process. The NPI-adoption fractions for CI, HQ, and SC are fixed during the simulation. The SD-adoption fraction, however, is chosen according to an optimised  assignment profile, presented in Table~\ref{tab:dynamic_SD}.

The resultant profile of SD-adoption is produced as a result of: (i) partitioning the simulated timeline with a number $h$ of change-points, limited by the number of the modelled pandemic phases, $h < 6$; (ii) varying these change-points in increments of 5 days; (iii) varying the fractions within each partitioned period, in increments of 0.1 between $SD_{min} = 0$ and $SD_{max} = 0.7$, with fractions $SD_{max} > 0.7$ assumed to be infeasible.

\renewcommand{\arraystretch}{1.5}
\begin{table}
    \centering
    \resizebox{\textwidth}{!}{
    \begin{tabular}{c|c|c|c}
        Parameter & Value & Distribution & Notes \\
        \hline \hline
         $\kappa$ & 23.0 & constant & global transmission scalar \\\hline
         $T_{inc}$ & 3 days (mean)  & lognormal ($\mu = 1.013,~\sigma = 0.413$)  & incubation period \\\hline
         $T_{rec}$ & 9 days (mean)  & uniform [7, 11]  & recovery period \\\hline
         $\alpha_{asymp}$ & 0.3  & constant  & asymptomatic transmission scalar \\\hline
         $\sigma_a$ & 0.67  & constant & probability of symptoms (age $<$ 18)  \\ \hline
         $\sigma_c$ & 0.268  & constant & probability of symptoms (age 18+)  \\ \hline
         $\pi_{symp}$ & 0.1  & constant & daily case detection probability (symptomatic)  \\ \hline
         $\pi_{asymp}$ & 0.01  & constant & daily case detection probability (asymptomatic)  \\ \hline
    \end{tabular}}
    \caption{Main input parameters for AMTraC-19 transmission model. }
    \label{tab:main_transmission}
\end{table}

\renewcommand{\arraystretch}{1.5}
\begin{table}
    \centering
    \resizebox{0.8\textwidth}{!}{
    \begin{tabular}{c|c|c|c}
        Parameter & Value and 95\% CI & Sample size  & Notes \\
        \hline \hline
         $R_0$ & 19.56 [19.12, 19.65] & 7,548 & basic reproductive ratio  \\\hline
         $T_{gen}$ & 5.42 [5.38, 5.44] & 7,548 & generation/serial interval  \\\hline
    \end{tabular}}
    \caption{Derived epidemiological parameters. }
    \label{tab:epi_parameters}
\end{table}

\renewcommand{\arraystretch}{1.3}
\begin{table}
    \centering
    \resizebox{\textwidth}{!}{
    \begin{tabular}{c|c|c|c|c|c|c|c}
         & \multicolumn{3}{c|}{Macro-distancing} & \multicolumn{4}{c}{Micro-distancing (interaction strengths)}\\
        \hline 
    Intervention & Compliance level & Duration $T$ & Threshold &    Household    & Community & Workplace \textbackslash School & Duration t  \\ \hline \hline
    CI & 0.7 & 196 & 0 & 1.0 & 0.25 & 0.25 & $D(i)$ \\ \hline
    HQ & 0.5 & 196 & 0 & 2.0 & 0.25 & 0.25 & 7 \\ \hline
    $SC^c$ & 1.0 & 110 & 100 & 1.0 &0.5 &0 & 110 \\ \hline
    $SC^a$ & 0.25 & 110 & 100 & 1.0 &0.5 &0 & 110 \\ \hline
    Static SD & $[0.2,0.7]$ & 196 & 400 & 1.0 &0.25 &0.1 & 196 \\ \hline
    \end{tabular}}
    \caption{The macro-distancing parameters and interaction strengths of NPIs in the studied scenarios. The micro-duration of CI is limited by the disease progression in the affected agent $i$,  $D(i)$. }
    \label{tab:NPI}
\end{table}

\renewcommand{\arraystretch}{1.5}
\begin{table}
    \centering
    \begin{tabular}{c|c|c|c}
    \hline
        \multicolumn{2}{c|}{Dynamic profile (23.4M agents)} & \multicolumn{2}{c}{Dynamic profile (scaled to 25.8M agents)} \\
        \hline
        SD-adoption fraction & Simulation period (days) &SD-adoption fraction & Simulation period (days) \\
        \hline \hline
         0.3 & 0-54 & 0.3 & 0-54  \\\hline
         0.7 & 55-94 & 0.7 & 55-94  \\ \hline
         0.6 & 95-109 & 0.6 & 95-109  \\ \hline
         0.4 & 110-124 & \textbf{0.5} & 110-\textbf{129} \\ \hline
         0.5 & 125-139  & \textbf{0.6} & \textbf{130}-139 \\ \hline
         0.2 & 140-196  & \textbf{0.3} & 140-196\\ \hline
    \end{tabular}
    \caption{Macro-distancing dynamic SD-adoption fractions optimised for (left): approximately 23.4M population (2016 census), and (right): approximately 25.8M population (scaled by 10\% relative to the 2016 census). The scaling-induced differences in best-fit SD-adoption fractions and simulation periods are highlighted in bold. Micro-distancing interaction strengths  are the same as in Table~\ref{tab:NPI}. Note that the initial SD-adoption of 0.3 is triggered when cumulative incidence reaches 400 around day 18 (specific days vary between different runs). }
    \label{tab:dynamic_SD}
\end{table}

\subsection{Vaccination modelling}
We simulated a pre-emptive vaccination rollout which immunised 10.53M (45\% of the population) with ``priority'' vaccine and 10.53M (45\% of the population) with ``general'' vaccine, reaching 90\% vaccination coverage nationwide. The coverage included 3.4M agents under 18 years of age ($age <18$), 14.3M agents between 18 and 65 years of age ($18 \leq age <65$), and 3.4M agents  at or over 65 years of age ($age \geq 65$), see Table~\ref{tab:vac_parameters}. In each simulation, the agents were immunised prior to the start of the Omicron stage. 

In setting vaccine efficacy levels, we followed study of Andrews et al.~\citep{Andrews2022} which reported that after 2-4 weeks, the efficacy of boosted BNT162b2 (Pfizer/BioNTech) is 67.2\% (95\% CI, 66.5 to 67.8) and the efficacy of  mRNA-1273 (Moderna) as  73.9\% (95\% CI, 73.1 to 74.6). For simplicity, we categorise these two vaccines as the ``priority'' vaccine with clinical efficacy set at $VE^c \approx 0.7$. Lower vaccine efficacy was reported among people who received ChAdOx1 nCoV-19 (Oxford/AstraZeneca), and we set $VE^c \approx 0.5$ for ``general'' vaccine. The clinical efficacy $VE^c$ is further split into the efficacy for susceptibility ($VE^s$) and the efficacy for disease ($VE^d$), following prior studies~\citep{zachreson2021how}:
\begin{equation}
    VE^c = VE^d + VE^s - VE^s \times VE^d
\end{equation}
where $VE^d = VE^s = 0.452$ for priority vaccine, and $VE^d = VE^s = 0.293$ for general vaccine.
Compared to the Delta variant, a lesser efficacy against transmission ($VE^t$) against Omicron has been reported \citep{jamanetworkopen.2022}, and we set $VE^t = 0.4$ for both types of vaccines considered in this study. A sensitivity analysis testing a range of $VE^t$ and $VE^c$ values was performed in prior studies~\citep{zachreson2021how,chang2022simulating}, showing that the model is robust to changes in the efficacy components.

For all vaccinated agents $j$ we set $VE^t_j = VE^t$, $VE^s_j = VE^s$ and $VE^d_j = VE^d$, and for all unvaccinated agents  $VE^t_j = VE^s_j = VE^d_j = 0$. The transmission probability of infecting a susceptible agent $i$ is derived as follows:
\begin{equation}
    p_i(n) =  1 - \prod_{g \in G_i(n)} \left [1 - (1 - VE^s_i) F_g(i) \left( 1 - \prod_{j \in A_g\backslash\{i\}} (1 - (1 - VE^t_j) F_g(j) \ p^g_{j \rightarrow i}(n))  \right) \right ]
		\label{eq-ve}
\end{equation}
The probability of becoming ill (symptomatic) is further affected by the efficacy against disease ($VE^d_i$): $p^d_i(n) = (1 - VE^d_i) \ \sigma_{a|c} \ p_i(n)$, given the fractions $\sigma_a$ and $\sigma_c$ of symptomatic adults and children, respectively.

\renewcommand{\arraystretch}{1.5}
\begin{table}
    \centering
    \begin{tabular}{c|c|c}
        Parameter  & Value & Reference \\\hline \hline
        Priority vaccine coverage & 10.53M & 45\% of total population\\  \hline
        General vaccine coverage & 10.53M  &  45\% of total population \\ \hline
        Vaccine allocation $[age <18]$  & 3.4M & 16.1\% of vaccinated population \\ \hline
        Vaccine allocation $[18 \leq age \leq 65]$ & 14.3M & 67.8\% of vaccinated population \\  \hline
        Vaccine allocation $[age \geq 65]$ & 3.4M & 16.1\% of vaccinated population \\ \hline
        Priority, $VE_d$ & 0.7 & \citep{Andrews2022} \\ \hline
        Priority, $VE_s$ & 0.452 & derived \\ \hline
        Priority, $VE_d$ & 0.452 & derived \\ \hline
        Priority, $VE_i$ & 0.4 & \citep{jamanetworkopen.2022} \\ \hline
        General, $VE_d$ & 0.5 & \citep{Andrews2022}  \\ \hline
        General, $VE_s$ & 0.293 & derived \\ \hline
        General, $VE_d$ & 0.293 & derived \\ \hline
        General, $VE_i$ & 0.4 & \citep{jamanetworkopen.2022} \\ \hline
    \end{tabular}
    \caption{Parameters of pre-emptive vaccination rollout  simulated with the ABM }
    \label{tab:vac_parameters}
\end{table}

\subsection{Mortality statistics}
\label{mort}
\begin{figure}
    \centering
    \includegraphics[width=\columnwidth,trim=2cm 1cm 3cm 1cm,clip]{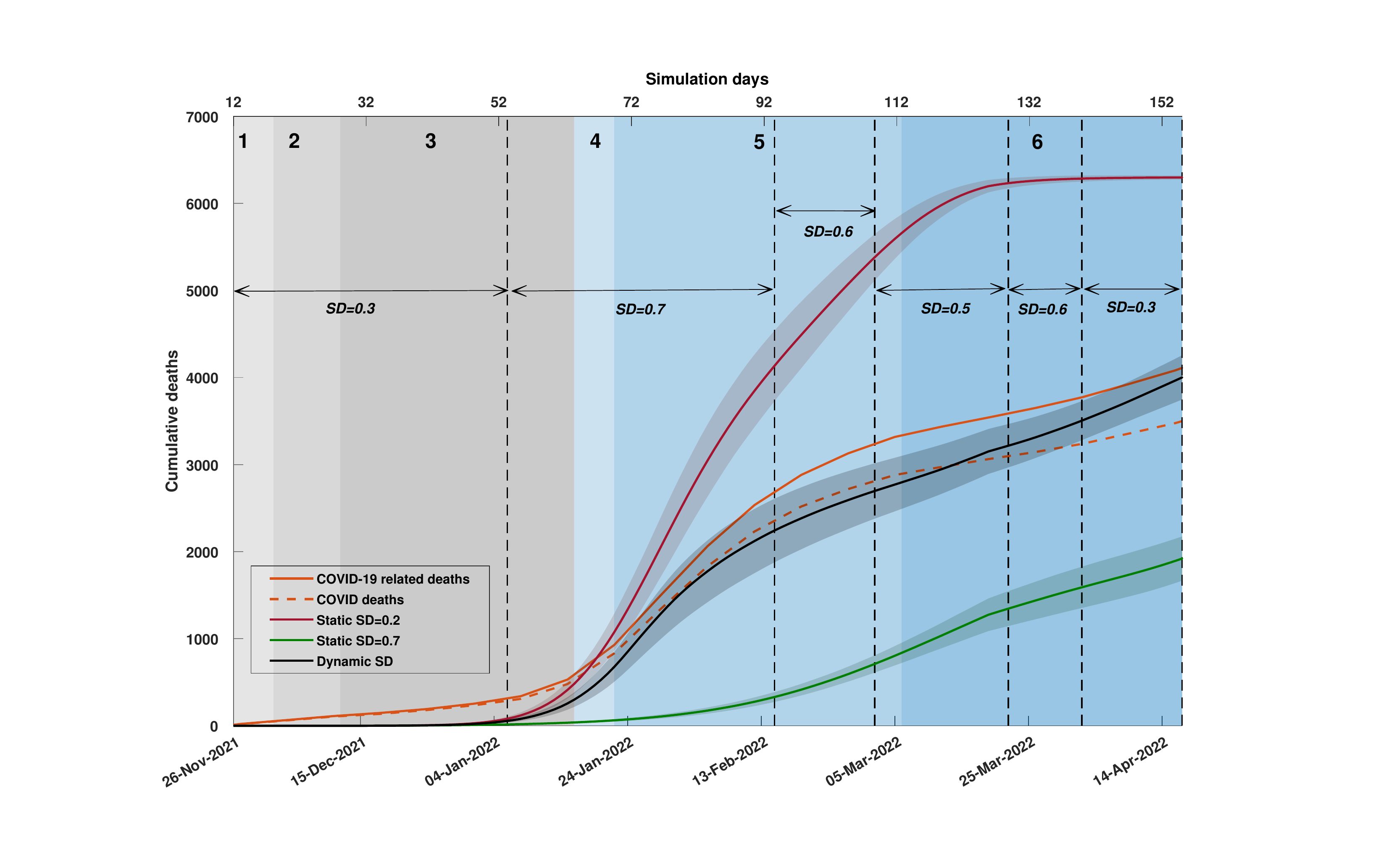}
    \caption{A comparison of cumulative deaths between dynamic social distancing (SD) levels (shown in solid black) and static SD levels ($SD_1 = 0.2$, shown in red; $SD_2 = 0.7$, shown in green). The simulated deaths are offset by 14 days. Coloured shaded areas around the solid line show standard deviation. Changes in dynamic SD-adoption are marked by vertical dashed black lines. Traces corresponding to each simulated scenario are computed as the average over 20 runs. SD adoption is combined with other interventions (i.e., school closures, case isolation, and home quarantine). Actual cumulative deaths are derived using reported weekly mortality (shown in orange; solid: COVID-19 related deaths; dashed: COVID-19 deaths). Shaded areas in grey and blue show the emergence of  variants of concern and sub-lineages over time, identified in weekly genomic surveillance reports (NSW Health). The timeline is divided into 6 phases as follows: 1) BA.1 detected, 2) Delta and BA.1 co-exist, 3) BA.1 dominant, 4) BA.2 detected, 5) BA.1 and BA.2 co-exist, and 6) BA.2 dominant.}
    \label{cum_death}
\end{figure}

Actual daily and cumulative deaths are derived from the reported weekly mortality for both (i) COVID-19  deaths, and (ii) COVID-19 related deaths, see Figures~\ref{actual}.c,~\ref{daily_death} and~\ref{cum_death}. This distinction differentiates between (i) doctor-certified deaths where COVID-19 is the underlying cause of death, and (ii) deaths where COVID-19 is both the underlying cause of death or a contributing factor (i.e., dying from or with COVID-19)~\cite{Mortality2022}. 

\begin{figure}
    \centering
    \includegraphics[width=\columnwidth,trim=2cm 1cm 3cm 1cm,clip]{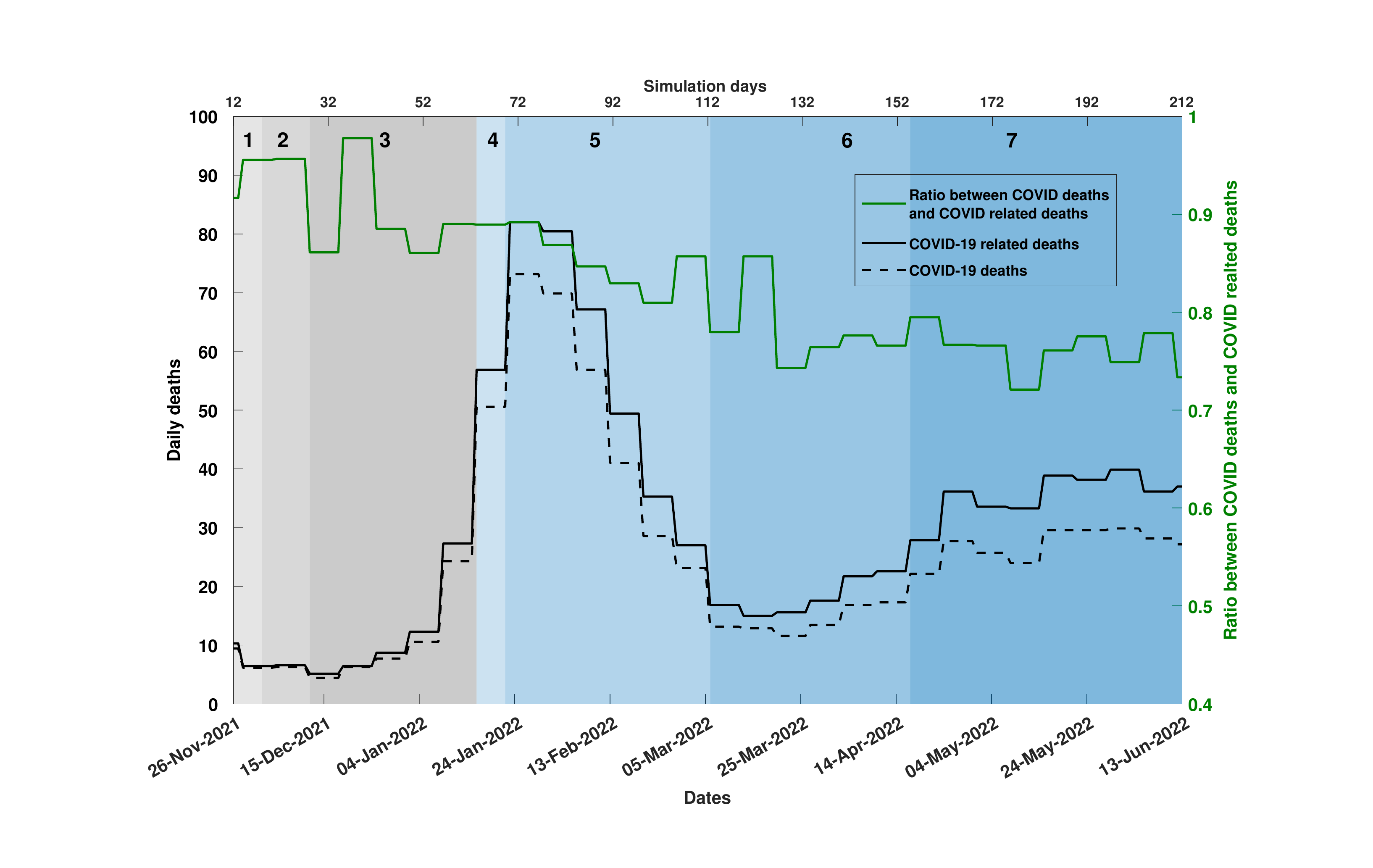}
    \caption{COVID-19 mortality data for Australia between 26th November 2021 and 13th June 2022. Shaded areas in grey and blue show the emergence of  variants of concern and sub-lineages over time, identified in weekly genomic surveillance reports (NSW Health). The timeline is divided into 7 phases as follows: 1) BA.1 detected, 2) Delta and BA.1 co-exist, 3) BA.1 dominant, 4) BA.2 detected, 5) BA.1 and BA.2 co-exist, 6) BA.2 dominant, and 7) BA.2, BA.4 and BA.5 co-exist. Solid black line (y-axis, left): COVID-19 related deaths; dashed black line (y-axis, left): COVID-19 deaths. Solid green line (y-axis, right): ratio between COVID-19 deaths and COVID-19 related deaths, with an average of 0.77 during phases 6 and 7.}
    \label{ratio_death}
\end{figure}

During phases 6 and 7 of the Omicron pandemic stage, these two causes of death diverged more markedly, with the ratio between COVID-19 deaths and COVID-19 related deaths averaging to $0.77$, see Fig.~\ref{ratio_death}. This ratio may also be used, during phases 6 and 7, as a proxy to differentiate between the ICU occupancy resulting from or with COVID-19, as shown in Fig.~\ref{ICU}, where the corresponding simulated trajectories are adjusted for phase 6 by scaling with $0.77$.

\subsection{Modelling disease burden}
\begin{figure}
    \centering
    \includegraphics[width=\columnwidth]{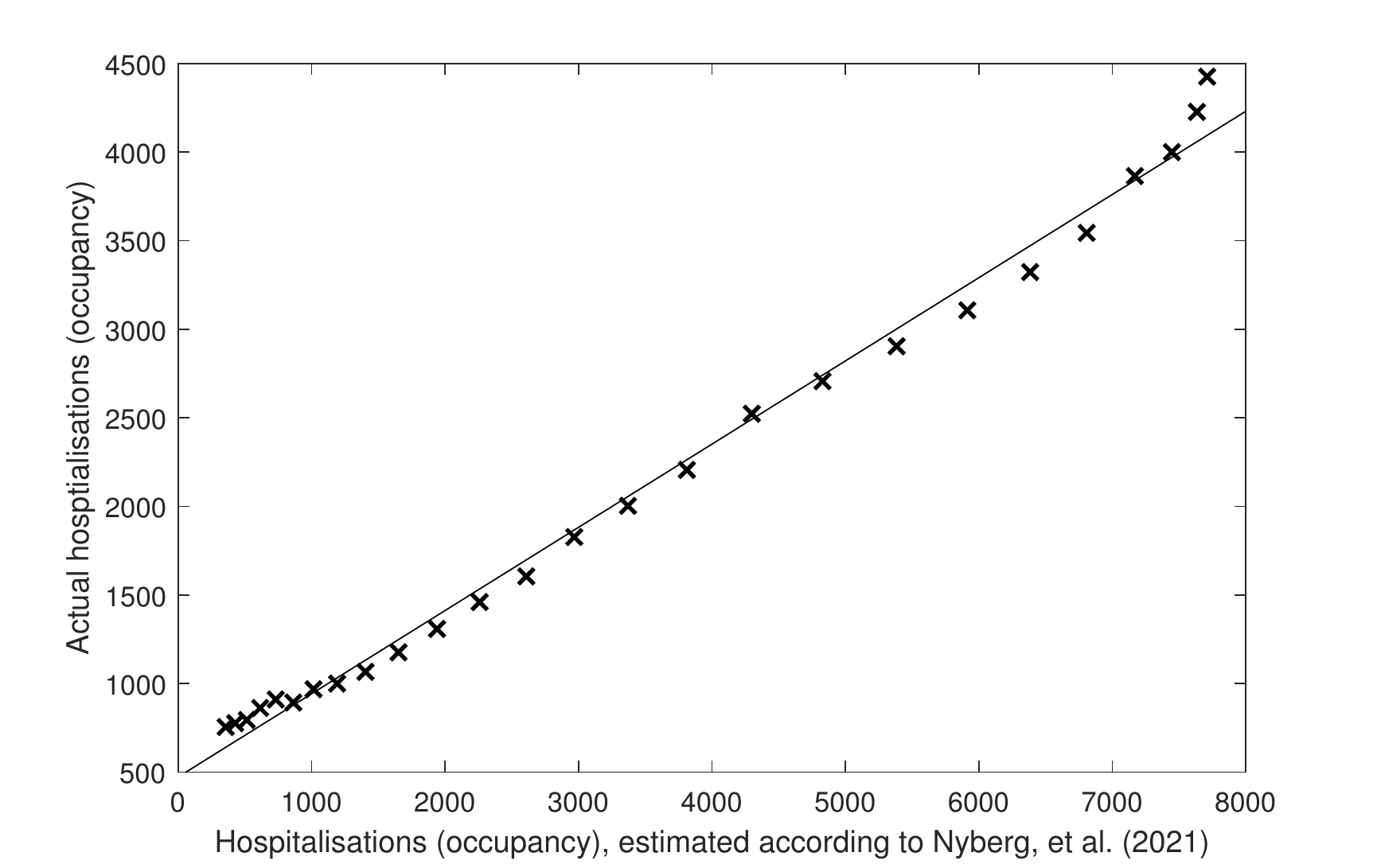}
    \caption{Linear regression between case hospitalisation risk (CHRs) for the Alpha variant and the hospitalisation cases reported in Australia between between 21 December 2021 and 15 January 2022. The regression produced a strong fit, with $R^2=0.9921$.  }
    \label{fig:linear_reg}
\end{figure}

We modelled disease burden in terms of hospitalisations (occupancy), ICU cases (occupancy), and daily and cumulative deaths. For daily hospitalisations, we scaled the age-dependent case hospitalisation risks (CHRs) from the Alpha variant (B.1.1.7)  reported by Nyberg et al.~\cite{nyberg2021risk} to the Omicron variant (BA.1)  by performing a linear regression between (i) the hospitalisation cases computed using CHRs for the Alpha variant, and (ii) the actual hospitalisation cases in Australia between 21 December 2021 and 15 January 2022. The regression shows a strong fit ($R^2=0.9921$) with the multiplier of $0.44$ and the additive constant of $472.9$ (Fig.~\ref{fig:linear_reg} and Table \ref{tab:clinical}). ospital admissions were set to follow infections by 7 days, with the offset derived by aligning the first peak for simulated and actual trajectories, and the average hospital stay was assumed to be 6 days, in agreement with various reports~\cite{JassatWaasila2022CsoC,iuliano2022trends,tobin2022hospital}. The risks of severe disease outcomes (i.e., hospitalisations, ICU admission and mortality) for vaccinated individuals are reduced by applying corresponding scalars summarised in Table~\ref{tab:vac_scalar}.

Daily ICU admissions were computed as fractions of the daily hospitalisations by applying age-dependent ratios. These ratios were approximated between the actual ICU occupancy and actual hospitalisations occupancy reported in NSW between 26 November 2021 and 12 February 2022 \citep{healthNSW2022}, summarised in Table \ref{tab:clinical}. ICU admissions were set to follow infections by 7 days, with the offset derived by aligning the first peak for simulated and actual trajectories, and the average ICU stay was assumed to be 5 days during the first five phases, and 4 days during phase 6~\cite{JassatWaasila2022CsoC,iuliano2022trends,tobin2022hospital}.

We then estimated the potential mortality, given the incidence cases, by scaling the age-dependent infection fatality rates (IFRs) for the Delta variant \citep{chang2022simulating}, reducing the risk of death for the Omicron variant to be 66\% lower compared to the Delta variant \citep{Lorenzo-Redondoo1806}.  Deaths were set to follow infections by 14 days, with the offset derived by aligning the first peak for simulated and actual trajectories.

In order to capture a  reduction in the disease severity of BA.2 relative to BA.1, reported elsewhere~\citep{Sievers2022, Wolter2022}, we  differentiated between sub-variants BA.1 and BA.2. The risks of hospitalisations, ICU admissions and mortality  for both vaccinated and unvaccinated individuals are  adjusted as follows:
\begin{equation}
  \begin{aligned}
    \text{CHR}_{\text{BA.k}} & = \text{CHR}_{\text{Omicron}} \times V^{\text{CHR}}_{\text{BA.k}} \\
    \text{ICU}_{\text{BA.k}} & = \text{ICU}_{\text{Omicron}} \times V^{\text{ICU}}_{\text{BA.k}} \\
    \text{IFR}_{\text{BA.k}} & = \text{IFR}_{\text{Omicron}} \times V^{\text{IFR}}_{\text{BA.k}} \\
  \end{aligned}
\end{equation}
where $k = 1$ or $k = 2$, and the default risks and rates for Omicron are presented in Table~\ref{tab:clinical}.

The vaccine efficacy against the severe disease caused by the Omicron variant is assumed to be 75\%~\citep{Nyberg2022}. This means that the age-dependent case hospitalisation risks for vaccinated individuals, $\text{CHR}^{vacc}_{\text{BA.k}}$, need to be further scaled down from the  values $\text{CHR}_{\text{BA.k}}$ for both sub-variants BA.1 and BA.2, by Omicron CHR vaccination scalar $V^{\text{CHR}}_{\text{Omicron}} = 0.25$:
\begin{equation}
    \text{CHR}^{vacc}_{\text{BA.k}} = \text{CHR}_{\text{BA.k}} \times V^{\text{CHR}}_{\text{Omicron}}
\end{equation}

The age-dependent risks of ICU admission for vaccinated individuals, $\text{ICU}^{vacc}_{\text{BA.k}}$, are also reduced from the risks $\text{ICU}_{\text{BA.k}}$ for both sub-variants, by  Omicron ICU vaccination scalar ($V^{\text{ICU}}_{\text{Omicron}}$): 
\begin{equation}
    \text{ICU}^{vacc}_{\text{BA.k}} = \text{ICU}_{\text{BA.k}} \times V^{\text{ICU}}_{\text{Omicron}}
\end{equation}
We varied parameter $V^{\text{ICU}}_{\text{Omicron}}$ in a range [0.5, 0.7] with 0.1 increment, and found that $V^{\text{ICU}}_{\text{Omicron}} = 0.6$ provides the best fit to the actual ICU occupancy in Australia.

The efficacy against death caused by the Omicron variant is assumed to be 80\%~\citep{Nyberg2022}. Thus, the age-dependent infection fatality rates for vaccinated individuals, $\text{IFR}^{vacc}_{\text{BA.k}}$,  are accordingly adjusted by  Omicron IFR vaccination scalar ($V^{\text{IFR}}_{\text{Omicron}} = 0.2$), reducing the  values of $\text{IFR}_{\text{BA.k}}$ for both sub-variants:
\begin{equation}
    \text{IFR}^{vacc}_{\text{BA.k}} = \text{IFR}_{\text{BA.k}} \times V^{\text{IFR}}_{\text{Omicron}}
\end{equation}

Table~\ref{tab:vac_scalar} summarises different vaccination scalars used for adjusting risks and rates for the Omicron variant and BA.1/BA.2 differentiation.

\renewcommand{\arraystretch}{1.5}
\begin{table}
    \centering
    \resizebox{\textwidth}{!}{
    \begin{tabular}{l|c|c|c|c|c|c|c|c|c}
        Rate \% \textbackslash \ Age & 0-9 & 10-19 & 20-29 & 30-39 & 40-49 & 50-59 & 60-69& 70-79 & 80+ \\
        \hline \hline
        $\text{CHR}_{\text{Alpha}}$ & 0.9 & 0.7 & 1.9 & 3.4 & 5.0 & 7.2 & 10.6 & 16.9 & 21.7    \\\hline
         $\text{CHR}_{\text{Omicron}}$ & 0.40 & 0.31 & 0.84 & 1.50 & 2.20 & 3.17 & 4.66 & 7.44 & 9.55 \\\hline
        $\text{IFR}_{\text{Delta}}$ & 0.0012 & 0.0024 & 0.010 & 0.011 & 0.036 & 0.11 & 0.68 & 1.81 & 4.38 \\ \hline
        $\text{IFR}_{\text{Omicron}}$ & 0.0004 & 0.0008 & 0.0034 & 0.0037 & 0.0122 & 0.0374 & 0.2312 & 0.6154 & 1.4892 \\ \hline
        $\text{ICU}_{\text{Omicron}}$ & 0.034 & 0.066 & 0.052 & 0.075 & 0.12 & 0.15 & 0.17 & 0.14 & 0.06 \\ \hline
    \end{tabular}}
    \caption{Estimates of age-dependent case hospitalisation risks (CHRs, \%), infection fatality rates (IFRs, \%) and ICU admission rates. ICU admission rates are published in weekly COVID-19 surveillance reports prepared by NSW Health \citep{healthNSW2022}. IFRs for Omicron are estimated to be 66\% lower than the Delta variant \cite{cdc2022}. }
    \label{tab:clinical}
\end{table}

\renewcommand{\arraystretch}{1.5}
\begin{table}
    \centering
    \begin{tabular}{l|c|c}
        Parameter & Value & Notes \\\hline \hline
         Omicron CHR vaccination scalar, $V^{\text{CHR}}_{\text{Omicron}}$  & 0.25 &  \citep{Nyberg2022}\\  \hline
         Omicron ICU vaccination scalar,  $V^{\text{ICU}}_{\text{Omicron}}$ & 0.60 & derived  \\ \hline
         Omicron IFR vaccination scalar, $V^{\text{IFR}}_{\text{Omicron}}$ & 0.20 & \citep{Nyberg2022} \\ \hline
         BA.1 CHR scalar, $V^{\text{CHR}}_{\text{BA.1}}$ & 1.0 & \citep{Sievers2022} \\  \hline
         BA.1 ICU scalar, $V^{\text{ICU}}_{\text{BA.1}}$ & 1.0 & \citep{Sievers2022} \\ \hline
         BA.1 IFR scalar, $V^{\text{IFR}}_{\text{BA.1}}$ & 1.0 & \citep{Sievers2022}  \\ \hline         
         BA.2 CHR scalar, $V^{\text{CHR}}_{\text{BA.2}}$ & 0.86 & \citep{Sievers2022} \\  \hline
         BA.2 ICU scalar, $V^{\text{ICU}}_{\text{BA.2}}$ & 0.85 & \citep{Sievers2022} \\ \hline
         BA.2 IFR scalar, $V^{\text{IFR}}_{\text{BA.2}}$ & 0.42 & \citep{Sievers2022}  \\ \hline
    \end{tabular}
    \caption{Vaccination scalars for the Omicron variant and BA.1/BA.2 differentiation. }
    \label{tab:vac_scalar}
\end{table}

\subsection{Calibration and sensitivity analysis}
\label{calibr-sens}
We calibrated the model to  transmission of the Omicron variant in Australia, focusing at sub-variant BA.1, but considering its mix with BA.2.  By varying the global transmission scalar ($\kappa$), we explored a range of the reproduction number ($R_0$). Our aim was to attain $R_0$ at least 3.1 times higher than the reproduction number of the Delta variant~\cite{obermeyer2022analysis}. The latter was calibrated in our prior studies, using similar ABMs, as being close to $R_0 = 6.20$ (with 95\% CI of 6.16--6.23, N = 6,609; at the scale of Australia)~\cite{chang2022simulating}, and $R_0 = 6.35$ (with 95\% CI 5.86--6.84, $N=300$; at the scale of New South Wales)~\cite{nguyen2022optimising}. 

To derive $R_0$ for the Omicron variant (as a mix of sub-variants BA.1 and BA.2), we carried out 7,548 simulations, each time randomly selecting an agent as the primary case, tracing transmissions, and counting only the direct secondary cases.  
We then used ``the attack rate pattern weighted index case'' method, eliminating a bias in selecting the primary case~\cite{germann2006mitigation,zachreson2020interfering}. The following age-specific attack rates were determined by the primary simulation: [0.0502, 0.1229, 0.1527, 0.4951, 0.1791],  for five age groups [0-4, 5-18, 19-29, 30-64, 65+]. The process produced $R_0 = 19.56$ (with 95\% CI 19.12--19.65, $N=7,548$), see Table~\ref{tab:epi_parameters}. 

We also varied the daily case detection probability (symptomatic) $\pi_{symp}$, in the range of 0.08 and 0.23, with $\pi_{symp} = 0.1$ producing the best fit to the actual incidence data. The parameters were calibrated to the Omicron stage in Australia between 26 November 2021 and 16 April 2022, as shown in Fig.~\ref{incidence} and summarised in Table~\ref{tab:main_transmission}. 

We performed a local point-based sensitivity analysis to examine the robustness of the ABM, quantifying the changes in the peak incidences in response to changes in parameters of interest, while using default values for the other input parameters, see Table~\ref{tab:main_transmission}.  We varied the following three parameters: the global transmission scalar ($\kappa$), the fraction of symptomatic cases ($\sigma_a$), and the infectivity  of asymptomatic cases ($\alpha_{asymp}$). The changes in these parameters were evaluated with respect to two incidence peaks (global maximum in mid-January, 2022; and local maximum in mid-April, 2022), shown in Figures~\ref{fig:sensi_kappa}--\ref{fig:sensi_inf}. 

\subsubsection{Global transmission scalar}
\label{sens-kappa}
\begin{figure}
    \centering
    \includegraphics[width=\columnwidth,trim=3cm 6cm 3cm 6cm,clip]{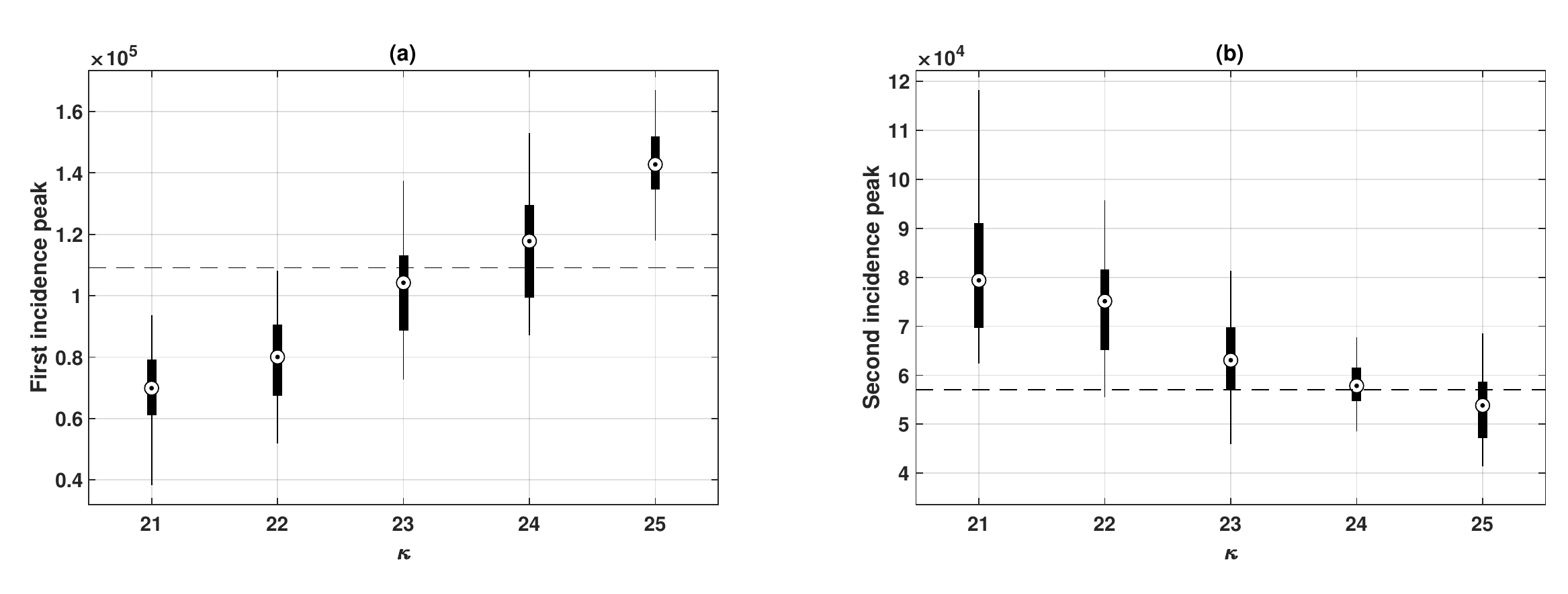}
    \caption{Local sensitivity analysis tracing incidence peaks with respect to changes in the global transmission scalar ($\kappa$) over 20 runs. The black dashed line marks the actual incidence peak. (a) First incidence peak (actual 7-day moving average incidence peak in early January). (b) Second incidence peak (actual 7-day moving average incidence peak in early April). }
    \label{fig:sensi_kappa}
\end{figure}

We varied the global transmission scalar ($\kappa$) in the range between 21 and 25, with the increment step of 1, centred around the default value at 23, simulated under the dynamic SD scenario. Other intervention conditions and parameters were unchanged. Fig.~\ref{fig:sensi_kappa} and Table \ref{tab:sensi_kappa} show the corresponding changes in the two incidence peaks.

In summary, higher $\kappa$ results in higher first peak and lower second peak. This is expected, since $\kappa$ directly scales the reproduction number $R_0$ from 17.81 to 20.75, as shown in Table \ref{tab:sensi_kappa}. In close proximity to the default value $\kappa=23$, the robustness of the model is not challenged by showing linear dependencies between the incidence peaks and $\kappa$. 

\renewcommand{\arraystretch}{1.3}
\begin{table}
    \centering
    \resizebox{\textwidth}{!}{
    \begin{tabular}{c|c|c|c|c|c|c|c|c}
         & \multicolumn{4}{c|}{First incidence peak} & \multicolumn{4}{|c}{Second incidence peak} \\ \hline 
         \hline
        $\kappa$ & mean & median & 25\% quantile & 75\% quantile  & mean & median & 25\% quantile & 75\% quantile \\ \hline
        21 ($R_0=17.81$)& 70,287 & 69,859   & 61,052 & 79,136   & 82,719 & 79,371 & 69,714 & 91,127 \\ \hline
        22 ($R_0=18.70$)& 79,781 & 80,025   & 67,438 & 90577   & 74,069 & 75,115 & 65,097 & 81,663 \\ \hline
 \textbf{23} (\boldmath $R_0=19.56$)&  \textbf{100,030} & \textbf{104,210} & \textbf{88,537} & \textbf{113,110}  & \textbf{64,033}  & \textbf{63,061}  & \textbf{57,124}  & \textbf{69,735} \\ \hline
        24 ($R_0=20.14$)& 116,890 & 117,770 & 99,451 & 129,660  & 57,488 & 57,833 & 54,070 & 61,579 \\ \hline
        25 ($R_0=20.75$)& 142,320 & 142,760 & 134,660 & 151,950 & 54,462 & 53,826 & 47,149 & 58,650 \\ \hline
    \end{tabular}}
    \caption{Statistics of the local sensitivity analysis tracing incidence peaks with respect to changes in the global transmission scalar ($\kappa$) over 20 runs. The default value is in bold. }
    \label{tab:sensi_kappa}
\end{table}

\subsubsection{Asymptomatic infectivity}
\label{sens-alpha}
\begin{figure}
    \centering
    \includegraphics[width=\columnwidth,trim=3cm 6cm 3cm 6cm,clip]{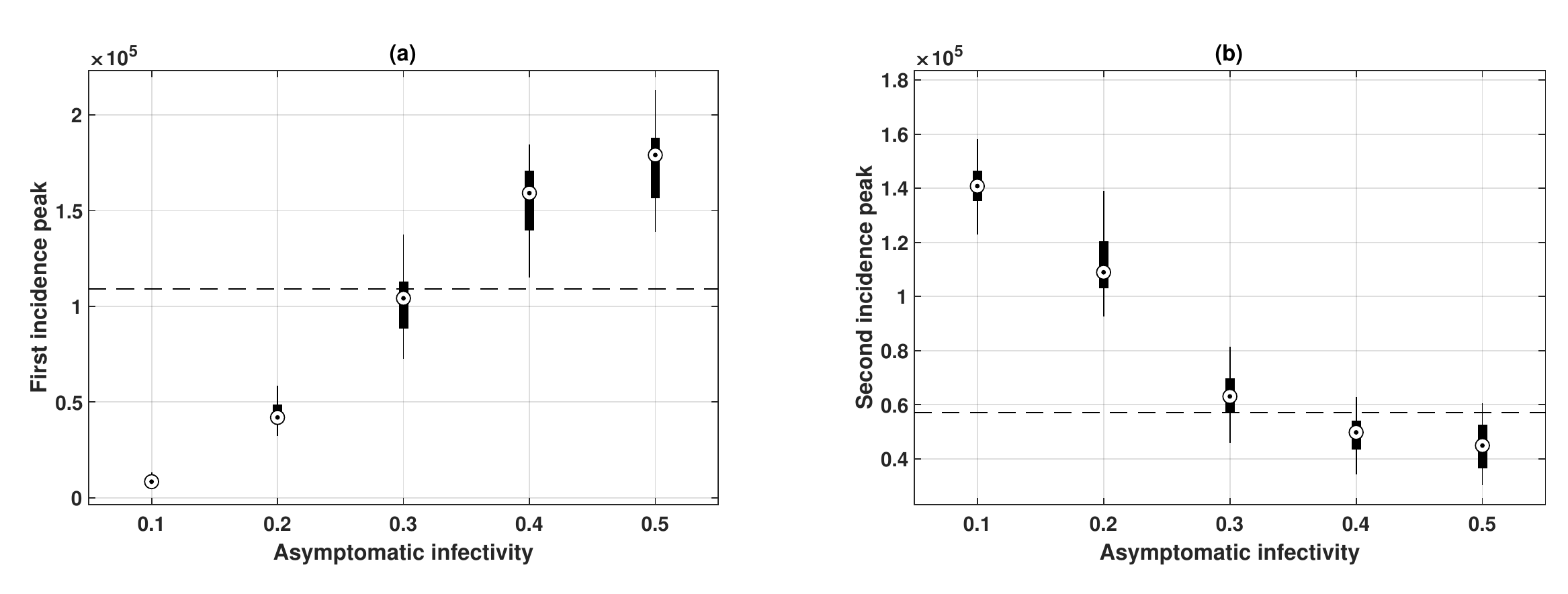}
    \caption{Local sensitivity analysis tracing incidence peaks with respect to changes in the infectivity  of asymptomatic cases ($\alpha_{asymp}$) over 20 runs. The black dashed line marks the actual incidence peak. (a) First incidence peak (actual 7-day moving average incidence peak in early January). (b) Second incidence peak (actual 7-day moving average incidence peak in early April). }
    \label{fig:sensi_inf}
\end{figure}

We model asymptomatic individuals with lower infectivity compared to their symptomatic counterparts. The asymptomatic infectivity is governed by asymptomatic infectivity $\alpha_{asymp}$ varying between 0 and 1, interpreted as the relative infectivity of an asymptomatic individual compared to the maximum level of an symptomatic case. We varied $\alpha_{asymp}$ within the range $[0.1, 0.5]$ with an increment step of 0.1. Fig.~\ref{fig:sensi_inf} and Table \ref{tab:sensi_inf} show the corresponding changes in two incidence peaks.

Higher asymptomatic infectivity produces higher first incidence peak and lower second incidence peak. At the lower bound $\alpha_{asymp}=0.1$, the first peak incidence has mean value of $8,892$ and increases to $171,770$ at the higher bound $\alpha_{asymp}=0.5$. The decrease in second incidence, on the other hand, is less sensitive to the change of $\alpha_{asymp}$. The default setting $\alpha_{asymp}=0.3$ is in concordance with the actual incidence peaks (marked by black dashed line). 

\renewcommand{\arraystretch}{1.3}
\begin{table}
    \centering
        \resizebox{\textwidth}{!}{
    \begin{tabular}{c|c|c|c|c|c|c|c|c}
         & \multicolumn{4}{c|}{First incidence peak} & \multicolumn{4}{|c}{Second incidence peak} \\ \hline 
         \hline
        $\alpha_{asymp}$ & mean & median & 25\% quantile & 75\% quantile  & mean & median & 25\% quantile & 75\% quantile \\ \hline
        0.1 & 8,892   & 8,403   & 7,356  & 9,989    & 141,470 & 140,790 & 135,290 & 146,450 \\ \hline
        0.2 & 44,296  & 41,952  & 40,036 & 48,736   & 111,090 & 108,940 & 103,010 & 120,390 \\ \hline
        \textbf{0.3} & \textbf{100,030} & \textbf{104,210} & \textbf{88,537} & \textbf{113,110}  & \textbf{64,033}  & \textbf{63,061}  & \textbf{57,124}  & \textbf{69,735} \\ \hline
        0.4 & 151,870 & 159,190 & 139,620 & 170,920 & 48,993  & 49,769  & 43,323  & 54,205 \\ \hline
        0.5 & 171,170 & 179,040 & 156,550 & 188,170 & 45,971  & 44,908  & 36,530  & 52,599 \\ \hline
    \end{tabular}}
    \caption{Statistics of the local sensitivity analysis tracing incidence peaks with respect to changes in the infectivity  of asymptomatic cases ($\alpha_{asymp}$), over 20 runs. The default value is in bold.}
    \label{tab:sensi_inf}
\end{table}

\subsubsection{Symptomatic fraction (adults)}
\label{sens-sigma}
\begin{figure}
    \centering
    \includegraphics[width=\columnwidth,trim=3cm 6cm 3cm 6cm,clip]{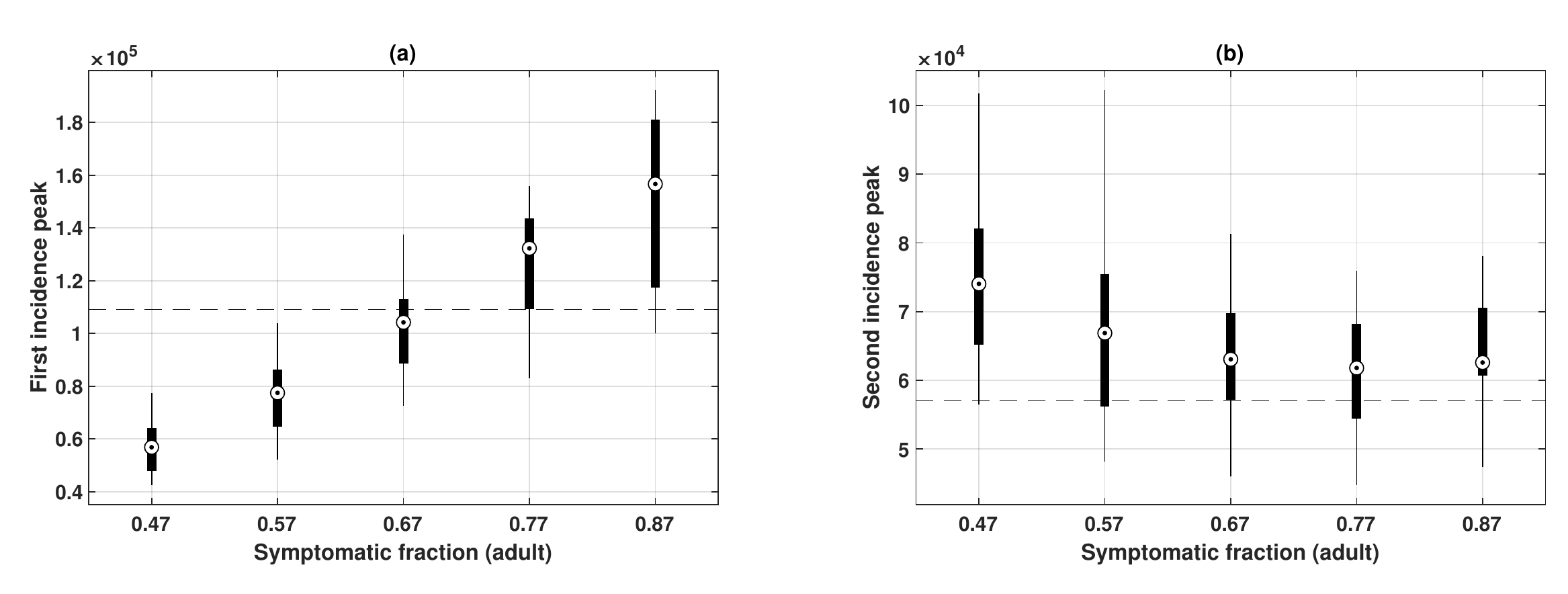}
    \caption{Local sensitivity analysis tracing incidence peaks with respect to changes in the fraction of symptomatic cases ($\sigma_a$) over 20 runs. The black dashed line marks the actual incidence peak. (a) First incidence peak (actual 7-day moving average incidence peak in early January). (b) Second incidence peak (actual 7-day moving average incidence peak in early April). }
    \label{fig:sensi_symp}
\end{figure}

Our model splits infected adults into symptomatic and asymptomatic cases, governed by the symptomatic fraction (adults), $\sigma_a$. This input parameter was tested within the range $[0.47, 0.87]$ with an incremental step of 0.1, centred around the default value of 0.67. Fig.~\ref{fig:sensi_symp} and Table \ref{tab:sensi_symp} summarise the changes in incidence peaks.

As $\sigma_a$ increases, the first incidence peak is more responsive to parameter change, increasing from 57,272 (mean) at the lower bound at $\sigma_a=0.47$ to 151,550 (mean) at the higher bound $\sigma_a=0.87$, while the second incidence shows low sensitivity with the peak reducing from 74,743 (mean) to 66,272 (mean).

\renewcommand{\arraystretch}{1.3}
\begin{table}
    \centering
    \resizebox{\textwidth}{!}{
    \begin{tabular}{c|c|c|c|c|c|c|c|c}
         & \multicolumn{4}{c|}{First incidence peak} & \multicolumn{4}{|c}{Second incidence peak} \\ \hline 
         \hline
        $\sigma_a$ & mean & median & 25\% quantile & 75\% quantile  & mean & median & 25\% quantile & 75\% quantile \\ \hline
         0.47 & 57,272 & 56,841 & 47,895 & 64,190 & 74,743 & 74,027 & 65,218 & 82,092 \\ \hline
         0.57 & 76,324 & 77,454 & 64,723 & 86,298 & 68,156 & 66,868 & 56,221 & 75,436 \\ \hline
         \textbf{0.67} &  \textbf{100,030} & \textbf{104,210} & \textbf{88,537} & \textbf{113,110}  & \textbf{64,033}  & \textbf{63,061}  & \textbf{57,124}  & \textbf{69,735} \\ \hline
         0.77 & 126,540 & 132,270 & 109,160 & 143,740 & 61,268 & 61,795 & 54,413 & 68,199 \\ \hline
         0.87 & 151,550 & 156,640 & 117,320 & 180,980 & 66,272 & 62,589 & 60,710 & 70,509 \\ \hline
    \end{tabular}}
    \caption{Statistics of the local sensitivity analysis tracing incidence peaks with respect to changes in the fraction of symptomatic cases ($\sigma_a$), over 20 runs. The default value is in bold.}
    \label{tab:sensi_symp}
\end{table}

\subsubsection{Summary}
We performed local sensitivity analysis by varying three key input parameters (while keeping the remaining parameters at their default values in dynamic SD scenario, using the population of 23.4M agents and the corresponding SD assignment shown in Table~\ref{tab:dynamic_SD}.Left): global transmission scalar ($\kappa$), the fraction of symptomatic cases ($\sigma_a$), and the infectivity  of asymptomatic cases ($\alpha_{asymp}$). We then tracked the changes in two incidence peaks, in order to measure the impact of these parameters. We observed that while both incidence peaks are sensitive to changes in infectivity of asymptomatic cases, the first incidence peak tends to be more sensitive to parameter change than the second incidence peak. Nevertheless, the model produces robust results for the parameter values in close proximity to the default values, which are in agreement with the actual incidence peaks.

\subsection{Google mobility data}
\begin{figure}
    \centering
    \includegraphics[width=\columnwidth,trim=1.5cm 1cm 3cm 1cm,clip]{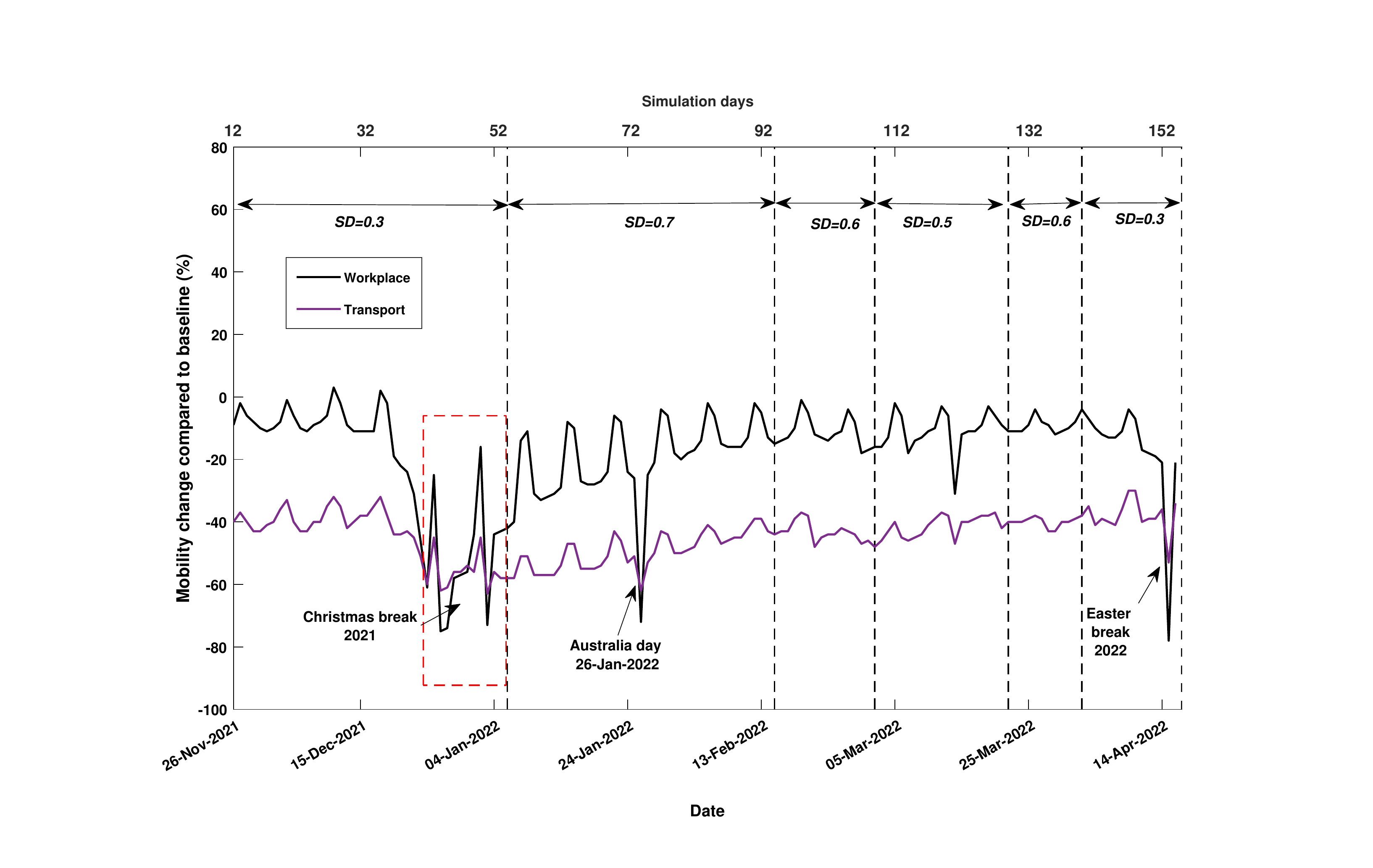}
    \caption{Movement trends between 26 November 2021 to 16 April 2022 in Australia at workplace and transport stations, compared to the baseline (i.e., the median value from the 5-week period between 3 January to 6 February 2020) \citep{Google}. The dashed black lines mark the ``retrodictively'' estimated SD periods corresponding to the simulation results shown in Fig. \ref{incidence}.}
    \label{fig:mobility}
\end{figure}

We use Google COVID-19 mobility reports between 26 November 2021 and 16 April 2022~\citep{Google} to explore qualitative agreement between mobility trends and the ''retrodictive'' SD levels and corresponding periods. Here, we use mobility trends observed in workplace and transit stations (i.e., transport) as a secondary proxy because these two categories are more likely to have significantly reduced social interactions during SD periods. We find satisfactory agreement between the ``retrodictive'' SD periods and Google mobility reports by observing greater reduction in mobility during the period with high SD (e.g., higher reduction during periods with $SD=0.7$ and $SD=0.6$ than periods with $SD=0.4$ and $SD=0.5$, as shown in Fig.~\ref{fig:mobility} and Table~\ref{tab:mobility}).

\renewcommand{\arraystretch}{1.3}
\begin{table}
    \centering    
    \resizebox{\textwidth}{!}{
    \begin{tabular}{c|c|c|c|c}
    Category  & 26/Nov/21 - 18/Dec/21 & 07/Jan/22 - 02/Mar/22 & 03/Mar/22 - 02/Apr/22 & 03/Apr/22 - 16/Apr/22 \\ 
              &  SD=0.3  & SD= 0.7 and 0.6 & SD=0.5 and 0.6 & SD=0.3 \\ \hline \hline
    Workplace (\%) & -6.96  & -17.07 & -10.06  & -12.67 \\ \hline
    Transport (\%) & -38.50 & -47.58 & -40.77 & -37.17 \\ \hline
    \end{tabular}}
    \caption{Mean mobility reduction (\%) during a period between 26 November 2021 and 16 April 2022 in Australia, at workplace and transport stations, compared to the baseline (i.e., the median value from the pre-pandemic 5-week period between 3 January to 6 February 2020) \citep{Google}. The mean mobility reduction is in qualitative accordance with the SD levels in the specified periods identified in Figure \ref{fig:mobility}. Public holidays (e.g., Christmas and New Year break, Australia Day and Easter break) are excluded. }
    \label{tab:mobility}
\end{table}

\subsection{Comparison between dynamic and static SD-adoption}
Tables~\ref{tab:inc_peak} and~\ref{tab:clinical_peak} summarise data for the two incidence peaks during the Omicron stage of the pandemic in Australia.  Three simulated scenarios (dynamic SD-adoption and two static SD-adoption alternatives) are contrasted with the actual data, in terms of daily incidence, hospitalisations, ICU occupancy and mortality. 

\renewcommand{\arraystretch}{1.3}
\begin{table}
    \centering
    \begin{tabular}{c|c|c|c|c}
        SD-adoption & 1st peak and STD & simulation day & 2nd peak and STD & simulation day  \\ \hline \hline
        0.2     & 185,000 (25,181) & 65 (6) & None & None\\ \hline
        0.7     & 41,000 (8,387) & 108 (7) & 56,000 (9,417) & 141 (20)\\ \hline
        Dynamic SD  & 104,000 (24,136) & 60 (2) & 69,000 (12,935) & 136 (4)\\ \hline
        Actual      & 109,100       & 10 January 2022 & 57,020 & 2 April 2022 \\ \hline
    \end{tabular}
    \caption{Mean and standard deviation (STD) of incidence peaks in three SD scenarios: static SD level of 0.2 and 0.7, and dynamic SD levels specified in Table \ref{tab:dynamic_SD}. Incidence peak magnitude (daily cases) is averaged over 20 runs rounded to the nearest '000. The lower static SD-adoption, $SD_1 = 0.2$, fails to produce the second peak between 26 November 2021 to 16 April 2022.}
    \label{tab:inc_peak}
\end{table}

\renewcommand{\arraystretch}{1.3}
\begin{table}
    \centering
    \resizebox{\textwidth}{!}{
    \begin{tabular}{c|c|c|c|c|c}
        SD-adoption & 1st peak hospitalisations & 1st peak ICU & 2nd peak hospitalisations & 2nd peak ICU & Cumulative deaths \\ \hline \hline
        0.2     & 8,700 (1,075) & 600 (70) & None & None & 6,300 (29)\\ \hline
        0.7     & 2,100 (343) & 210 (22) & None & None & 2,000 (214)\\ \hline
        Dynamic SD  & 5,200 (1,015) & 370 (66) & 3000 (453) & 240 (29) & 4,200 (116)\\ \hline
        Actual  &   5,371 & 424 & 3,147 & Not applicable & 4,083 \\ \hline 
    \end{tabular}}
    \caption{Mean and standard deviation (STD) of peak estimates of the hospitalisations and ICU occupancy, and cumulative deaths (16 April 2022) in three SD scenarios: static SD level of 0.2 and 0.7, and dynamic SD levels specified in Table \ref{tab:dynamic_SD}. Peak magnitudes are averaged over 20 runs rounded to the nearest '00. The static SD-adoption fractions, $SD_1 = 0.2$ and $SD_2 = 0.7$, fail to produce the second peak within the considered timeline between 26 Nov 2021 to 14 Apr 2022.}
    \label{tab:clinical_peak}
\end{table}


\end{document}